\documentclass[aps,prd,nofootinbib,showpacs]{revtex4}
\usepackage{amssymb}
\usepackage{amsmath}
\usepackage{amsfonts}
\usepackage{epsfig}
\usepackage{appendix}
\def\uu{\langle \bar u u \rangle}
\def\dd{\langle \bar d d \rangle}

\newcommand{\seq}{\begin{subequations}}
\newcommand{\sen}{\end{subequations}}
\newcommand{\eq}{\begin{eqnarray}}
\newcommand{\en}{\end{eqnarray}}

\begin{document}

\title{  \bf  Flavor Changing Neutral Currents  Transition of the $\Sigma_{Q}$ to Nucleon   in Full QCD and Heavy Quark Effective Theory}
\author{ K. Azizi$^{1,\dag}$,  M. Bayar$^{2,\ddag}$,   M. T. Zeyrek$^{3,*}$, \\
 $^\dag$ Physics Division, Faculty of Arts and
Sciences, Do\u gu\c s University,
 Ac{\i}badem-Kad{\i}k\"oy,\\ 34722 Istanbul, Turkey\\
$^\ddag$ Department of Physics, Kocaeli University, 41380 Izmit, Turkey\\
$^*$ Physics Department, Middle East Technical University, 06531, Ankara, Turkey\\
$^1$kazizi@dogus.edu.tr \\
$^2$melahat.bayar@kocaeli.edu.tr\\
$^3$zeyrek@metu.edu.tr }

\begin{abstract}
The loop level flavor changing neutral currents transitions of the
$\Sigma_{b}\rightarrow n~l^+l^-$ and $\Sigma_{c}\rightarrow
p~l^+l^-$
 are investigated in full QCD and heavy quark effective theory in
the  light cone QCD sum rules approach. Using the most general form
of the   interpolating current for $\Sigma_{Q}$,   $Q=b$ or $c$, as
members of the recently discovered sextet heavy baryons with spin
1/2 and containing one heavy quark, the transition form factors are
calculated using two sets of input parameters entering the nucleon
distribution amplitudes. The obtained results are used to estimate the decay rates of
the corresponding transitions. Since such type transitions occurred
at loop level in the standard model,  they can be considered as good
candidates to search for the new physics effects beyond the SM.
\end{abstract}


\maketitle

\section{Introduction}
The $\Sigma_{b}\rightarrow n~l^+l^-$ and $\Sigma_{c}\rightarrow
p~l^+l^-$ are governed by flavor changing neutral currents (FCNC)
transitions of $b\rightarrow d$ and  $c\rightarrow u$, respectively.
These transitions are described via  electroweak penguin and weak
box diagrams in the standard model (SM) and they are sensitive to
new physics  contributing to penguin operators. Looking  for SUSY
particles \cite{susy}, light dark matter \cite{darkmatter} and also
probable fourth generation of the quarks is possible by
investigating such loop level transitions. This transitions  are
also good framework to reliable determination of the $V_{tb}$,
$V_{td}$, $V_{cb}$, and $V_{bu}$ as members of the
Cabibbo-Kobayashi-Maskawa (CKM) matrix,  CP and T violations and
polarization asymmetries. The $\Sigma_{b,c}$ as members of the spin
1/2 sextet heavy baryons containing a single heavy bottom or charm
quark are considered by their most general interpolating currents
which generalize the Ioffe current for these baryons. In the recent
years, important experimental progresses has been made in the
spectroscopy of the heavy baryons containing heavy b or c quark
\cite{Mattson,Ocherashvili,Acosta,Chistov,Aubert1,Abazov1,Aaltonen1,Solovieva}.
Having  the heavy quark makes these  states be  experimentally
narrow, so their isolation and detection are easy comparing with the
light baryons.
 Experimentally, investigation of the  semileptonic decays of the heavy
 baryons,   may be considered at large hadron collider (LHC) in the future,
 hence theoretical calculations of the decay properties can play  crucial role  in this respect.

In our two recent works, we analyzed the tree level semileptonic
decays of $\Sigma_b$ to proton \cite{kmayprd} and
$\Lambda_b(\Lambda_c)\rightarrow p(n)l\nu$  \cite{kmyh} in light
cone QCD sum rules. In full theory, these tree level transitions in
the SM are described  via six form factors (for details and more
about the works devoted to the semileptonic decays of the heavy
baryons using different phenomenological methods see
\cite{kmayprd,kmyh} and references therein). In the present work,
considering the long and short distance effects, we calculate the 12
form factors entering the semileptonic loop level
$\Sigma_{b}\rightarrow n~l^+l^-$ and $\Sigma_{c}\rightarrow
p~l^+l^-$ transitions using the light cone QCD sum rules in full
theory as well as heavy quark effective theory (HQET). The short
distance effects are calculated using the perturbation theory  and
long distance contributions are  expanded in terms of the nucleon
distribution amplitudes (DA's) with increasing twists near the light
cone, $x^2\simeq0$. We use the value of the eight independent
parameters entering to the nucleon DA's from two different sources:
predicted using a simple
model in which the deviation from the asymptotic
DAs is taken to be 1/3 of that suggested by
the QCD sum rule estimates
 \cite{Lenz} and obtained via lattice QCD
\cite{Gockeler1,Gockeler2,QCDSF}.  Using  the obtained form factors,
we predict the corresponding transition rates. Investigation of
these decays can also give essential information about the internal
structure of $\Sigma_{b,c}$ baryons as well as the nucleon DA's.

The layout  of the paper is as follows: in section II, we introduce
the theoretical framework to calculate   the form factors in light
cone QCD sum rules method in full theory. The HQET relations among
the form factors are also introduced in this section.  Section III
is devoted to the numerical analysis of the form factors and their
extrapolation in terms of the transferred momentum squired, $q^2$,
their HQET limit and our predictions for the decay rates obtained in
two different sets of parameters entering the nucleon distribution amplitudes.

\section{ Light Cone QCD Sum Rules for transition form factors}
Form factors play essential role in analyzing the
$\Sigma_{b}\rightarrow n~l^+l^-$ and $\Sigma_{c}\rightarrow
p~l^+l^-$ transitions.  At quark level, these decays proceed by loop
$b~(c)\rightarrow d~(u)$ transition, which  can be described by the
following effective Hamiltonian:
\begin{eqnarray}
{\cal H}_{eff} = \frac{G_F~\alpha V_{Q'Q}~V_{Q'q}^{^{*}}
}{2\sqrt2~\pi} \left\{\vphantom{\int_0^{x_2}}  C_{9}^{eff}~ \bar{q}
\gamma_\mu (1-\gamma_5) Q \bar l \gamma^\mu
 l +C_{10} ~\bar{q}
\gamma_\mu (1-\gamma_5) Q \bar l \gamma^\mu \gamma_{5}l-2
m_{Q}~C_{7}\frac{1}{q^{2}} ~\bar{q} i \sigma_{\mu\nu}q^{\nu}
(1+\gamma_5) Q \bar l \gamma^\mu l \right \},\nonumber\\
\end{eqnarray}
 where, $Q'$ refers to the $u,~c,~t$ for bottom case and $d,~s~,b$ for charm case, respectively.
 The main contributions come from the heavy quarks, so we will consider $Q'=t$ and $Q'=b$ respectively
  for the
$\Sigma_{b}\rightarrow n~l^+l^-$ and $\Sigma_{c}\rightarrow
p~l^+l^-$ transitions. The amplitude of the considered transitions
can be obtained by sandwiching  the above Hamiltonian between the
initial and final states. To proceed, we need to know the   the
matrix elements $\langle N \vert  J^{tr,I}_{\mu} \vert \Sigma_{Q}
\rangle $ and $\langle N \vert  J^{tr,II}_{\mu} \vert \Sigma_{Q}
\rangle $, where $J^{tr,I}_{\mu}(x)=\bar q(x) \gamma_\mu
(1-\gamma_5) Q(x)$ and $J^{tr,II}_{\mu}(x)=\bar{q}(x) i
\sigma_{\mu\nu}q^{\nu} (1+\gamma_5) Q(x)$ are transition currents
entering to the Hamiltonian. From the general philosophy of the QCD
sum rules, to obtain sum rules for the physical quantities we start
considering the following  correlation functions:
\begin{eqnarray}\label{T}
\Pi^I_{\mu}(p,q)&=&i\int d^{4}xe^{iqx}\langle N(p)\mid
T\{J^{tr,I}_{\mu}(x)\bar J^{\Sigma_{Q}}(0) \}\mid 0\rangle,\nonumber\\
\Pi^{II}_{\mu}(p,q)&=&i\int d^{4}xe^{iqx}\langle N(p)\mid
T\{J^{tr,II}_{\mu}(x)\bar J^{\Sigma_{Q}}(0) \}\mid 0\rangle,
\end{eqnarray}
where, $J^{\Sigma_{Q}}$ is interpolating currents of $\Sigma_{b(c)}$
baryon and $p$ denotes the proton (neutron) momentum and $q=(p+q)-p$
is the transferred momentum.  The main idea in QCD sum rules is to calculate the aforementioned correlation functions in two different ways:
\begin{itemize}
 \item  In theoretical side, the time ordering product of the initial state and transition current
 is expanded  in terms of nucleon distribution amplitudes  having different twists
 via the operator product expansion (OPE) at deep Euclidean region. By OPE the short and large distance
 effects are separated. The short distance contribution is calculated using the perturbation theory,
 while the long distance phenomena are parameterized in terms of nucleon DA's.
\item From phenomenological or physical side, they are calculated in terms of the hadronic parameters
via saturating them with a tower of hadrons with the same quantum numbers as the interpolating currents.
\end{itemize}

To get the sum rules for the physical quantities, the two above representations of the correlation functions are equated through the dispersion relation. To suppress the contribution of the higher states and continuum and isolate the ground state, the Borel transformation as well as continuum subtraction through quark-hadron duality assumption are applied to both sides of the sum rules expressions.

The first task is to calculate the aforementioned correlation
function from QCD side in deep Euclidean region where $(p+q)^2\ll0$.
To proceed, the explicit expression of the interpolating field of
the $\Sigma_{Q}$ baryon is needed. Considering the quantum numbers,
the most general form of interpolating current creating the
$\Sigma_Q$ from the vacuum can be written as

\begin{eqnarray}\label{cur.N}
J^{\Sigma_b}(x)&=&
\frac{-1}{\sqrt{2}}\varepsilon^{abc}\left[\vphantom{\int_0^{x_2}}\left\{\vphantom{\int_0^{x_2}}q_{1}^{T
a} (x)C Q^{b} (x) \right\}\gamma_5 q_{2}^{c}
(x)-\left\{\vphantom{\int_0^{x_2}}Q^{T a} (x)C q_{2}^{b} (x)
\right\}\gamma_5
q_{1}^{c} (x)\right. \nonumber \\
&& \left. +\beta\left\{\vphantom{\int_0^{x_2}}\{q_{1}^{T a} (x)C
\gamma_5 Q^{b} (x) \} q_{2}^{c} (x)-\{Q^{T a} (x)C \gamma_5
q_{2}^{b} (x) \} q_{1}^{c}
(x)\right\}\vphantom{\int_0^{x_2}}\right],
\end{eqnarray}
where, $C$ is the charge conjugation
operator and $\beta$ is an arbitrary parameter with $\beta=-1$
corresponding to the Ioffe current, $q_{1}$ and $q_{2}$ are the $u$ and $d$ quarks, respectively
 and $a,~b,~c$ are the color indices. Using the transition currents,
 and $J^{\Sigma_{Q}}
$and contracting out all quark pairs via the Wick's theorem, we
obtain the following representations of the correlation functions in
QCD side:
\begin{eqnarray}\label{mut.m}
\Pi^I_{\mu} &=& \frac{-i}{\sqrt{2}} \epsilon^{abc}\int d^4x e^{iqx}
\Bigg\{\Bigg(\Big[( C )_{\beta\eta} (\gamma_5)_{\rho\phi}-( C
)_{\phi\beta} (\gamma_5)_{\rho\eta}\Big]  +\beta\Bigg[(C \gamma_5
)_{\beta\eta}(I)_{\rho\phi}
 \nonumber \\
&-& (C \gamma_5 )_{\phi\beta}(I)_{\rho\eta} \Bigg]\Bigg) \Big[
\gamma_{\mu}(1-\gamma_5)
\Big]_{\sigma\theta}\Bigg\}S_Q(-x)_{\beta\sigma} \langle  N (p) |
\bar d_\eta^a(0)
\bar d_\theta^b(x)  \bar u_\phi^c(0) | 0\rangle ,\nonumber\\
\end{eqnarray}
\begin{eqnarray}\label{mut.mm}
\Pi^{II}_{\mu} &=& \frac{-i}{\sqrt{2}} \epsilon^{abc}\int d^4x e^{iqx}
\Bigg\{\Bigg(\Big[( C )_{\beta\eta} (\gamma_5)_{\rho\phi}-( C
)_{\phi\beta} (\gamma_5)_{\rho\eta}\Big]  +\beta\Bigg[(C \gamma_5
)_{\beta\eta}(I)_{\rho\phi}
 \nonumber \\
&-& (C \gamma_5 )_{\phi\beta}(I)_{\rho\eta} \Bigg]\Bigg) \Big[
i\sigma_{\mu\nu}q^\nu(1+\gamma_5)
\Big]_{\sigma\theta}\Bigg\}S_Q(-x)_{\beta\sigma} \langle  N (p) |
\bar d_\eta^a(0)
\bar d_\theta^b(x)  \bar u_\phi^c(0) | 0\rangle ,\nonumber\\
\end{eqnarray}
where, $ S_Q(x)$ is the heavy quark propagator and its expression is
given as \cite{Balitsky}:

\begin{eqnarray}\label{heavylightguy}
 S_Q (x)& =&  S_Q^{free} (x) - i g_s \int \frac{d^4 k}{(2\pi)^4}
e^{-ikx} \int_0^1 dv \Bigg[\frac{\not\!k + m_Q}{( m_Q^2-k^2)^2}
G^{\mu\nu}(vx) \sigma_{\mu\nu} + \frac{1}{m_Q^2-k^2} v x_\mu
G^{\mu\nu} \gamma_\nu \Bigg].
 \end{eqnarray}
where
\begin{eqnarray}\label{freeprop}
S^{free}_{Q}
&=&\frac{m_{Q}^{2}}{4\pi^{2}}\frac{K_{1}(m_{Q}\sqrt{-x^2})}{\sqrt{-x^2}}-i
\frac{m_{Q}^{2}\not\!x}{4\pi^{2}x^2}K_{2}(m_{Q}\sqrt{-x^2}),\nonumber\\
\end{eqnarray}
and  $K_i$ are the Bessel functions. When doing calculations, we
neglect the terms proportional to the gluon field strength tensor
because they are contributed mainly to the  four and five particle
distribution functions and expected to be very small in our case
\cite{17,18,Braun1b}. The matrix element $\langle  N (p)\mid
\epsilon^{abc}\bar d_{\eta}^{a}(0)\bar d_{\theta}^{b}(x)\bar
u_{\phi}^{c}(0)\mid 0\rangle$  appearing in Eqs. (\ref{mut.m},\ref{mut.mm})
denotes  the nucleon wave function, which is given in terms of some
calligraphic functions \cite{Lenz,17,18,Braun1b,8}:

\begin{eqnarray}\label{wave func}
&&4\langle0|\epsilon^{abc}d_\alpha^a(a_1 x)d_\beta^b(a_2
x)u_\gamma^c(a_3 x)|N(p)\rangle\nonumber\\
&=&\mathcal{S}_1m_{N}C_{\alpha\beta}(\gamma_5N)_{\gamma}+
\mathcal{S}_2m_{N}^2C_{\alpha\beta}(\rlap/x\gamma_5N)_{\gamma}\nonumber\\
&+& \mathcal{P}_1m_{N}(\gamma_5C)_{\alpha\beta}N_{\gamma}+
\mathcal{P}_2m_{N}^2(\gamma_5C)_{\alpha\beta}(\rlap/xN)_{\gamma}+
(\mathcal{V}_1+\frac{x^2m_{N}^2}{4}\mathcal{V}_1^M)(\rlap/pC)_{\alpha\beta}(\gamma_5N)_{\gamma}
\nonumber\\&+&
\mathcal{V}_2m_{N}(\rlap/pC)_{\alpha\beta}(\rlap/x\gamma_5N)_{\gamma}+
\mathcal{V}_3m_{N}(\gamma_\mu
C)_{\alpha\beta}(\gamma^\mu\gamma_5N)_{\gamma}+
\mathcal{V}_4m_{N}^2(\rlap/xC)_{\alpha\beta}(\gamma_5N)_{\gamma}\nonumber\\&+&
\mathcal{V}_5m_{N}^2(\gamma_\mu
C)_{\alpha\beta}(i\sigma^{\mu\nu}x_\nu\gamma_5N)_{\gamma} +
\mathcal{V}_6m_{N}^3(\rlap/xC)_{\alpha\beta}(\rlap/x\gamma_5N)_{\gamma}
+(\mathcal{A}_1\nonumber\\
&+&\frac{x^2m_{N}^2}{4}\mathcal{A}_1^M)(\rlap/p\gamma_5
C)_{\alpha\beta}N_{\gamma}+
\mathcal{A}_2m_{N}(\rlap/p\gamma_5C)_{\alpha\beta}(\rlap/xN)_{\gamma}+
\mathcal{A}_3m_{N}(\gamma_\mu\gamma_5 C)_{\alpha\beta}(\gamma^\mu
N)_{\gamma}\nonumber\\&+&
\mathcal{A}_4m_{N}^2(\rlap/x\gamma_5C)_{\alpha\beta}N_{\gamma}+
\mathcal{A}_5m_{N}^2(\gamma_\mu\gamma_5
C)_{\alpha\beta}(i\sigma^{\mu\nu}x_\nu N)_{\gamma}+
\mathcal{A}_6m_{N}^3(\rlap/x\gamma_5C)_{\alpha\beta}(\rlap/x
N)_{\gamma}\nonumber\\&+&(\mathcal{T}_1+\frac{x^2m_{N}^2}{4}\mathcal{T}_1^M)(p^\nu
i\sigma_{\mu\nu}C)_{\alpha\beta}(\gamma^\mu\gamma_5
N)_{\gamma}+\mathcal{T}_2m_{N}(x^\mu p^\nu
i\sigma_{\mu\nu}C)_{\alpha\beta}(\gamma_5 N)_{\gamma}\nonumber\\&+&
\mathcal{T}_3m_{N}(\sigma_{\mu\nu}C)_{\alpha\beta}(\sigma^{\mu\nu}\gamma_5
N)_{\gamma}+
\mathcal{T}_4m_{N}(p^\nu\sigma_{\mu\nu}C)_{\alpha\beta}(\sigma^{\mu\rho}x_\rho\gamma_5
N)_{\gamma}\nonumber\\&+& \mathcal{T}_5m_{N}^2(x^\nu
i\sigma_{\mu\nu}C)_{\alpha\beta}(\gamma^\mu\gamma_5 N)_{\gamma}+
\mathcal{T}_6m_{N}^2(x^\mu p^\nu
i\sigma_{\mu\nu}C)_{\alpha\beta}(\rlap/x\gamma_5
N)_{\gamma}\nonumber\\&+&
\mathcal{T}_7m_{N}^2(\sigma_{\mu\nu}C)_{\alpha\beta}(\sigma^{\mu\nu}\rlap/x\gamma_5
N)_{\gamma}+
\mathcal{T}_8m_{N}^3(x^\nu\sigma_{\mu\nu}C)_{\alpha\beta}(\sigma^{\mu\rho}x_\rho\gamma_5
N)_{\gamma}.
\end{eqnarray}
The calligraphic functions  have not definite twists but they can be
expressed in terms of the nucleon distribution amplitudes (DA's)
with definite and  increasing twists by the help of   the scalar
product $px$ and the parameters $a_i$, $i=1,2,3$. The explicit
expressions for  scalar, pseudo-scalar, vector, axial vector and
tensor DA's for nucleons are given in Tables \ref{tab:1},
\ref{tab:2}, \ref{tab:3}, \ref{tab:4} and \ref{tab:5}, respectively.
\begin{table}[h]
\centering
\begin{tabular}{|c|} \hline
$\mathcal{S}_1 = S_1$\\\cline{1-1}\hline
 $2px\mathcal{S}_2=S_1-S_2$ \\\cline{1-1}
   \end{tabular}
\vspace{0.3cm} \caption{Relations between the calligraphic functions
and nucleon scalar DA's.}\label{tab:1}
\end{table}
\begin{table}[h]
\centering
\begin{tabular}{|c|} \hline
  $\mathcal{P}_1=P_1$\\\cline{1-1}
  $2px\mathcal{P}_2=P_1-P_2$ \\\cline{1-1}
   \end{tabular}
\vspace{0.3cm} \caption{Relations between the calligraphic functions
and nucleon pseudo-scalar DA's.}\label{tab:2}
\end{table}
\begin{table}[h]
\centering
\begin{tabular}{|c|} \hline
  $\mathcal{V}_1=V_1$ \\\cline{1-1}
  $2px\mathcal{V}_2=V_1-V_2-V_3$ \\\cline{1-1}
  $2\mathcal{V}_3=V_3$ \\\cline{1-1}
  $4px\mathcal{V}_4=-2V_1+V_3+V_4+2V_5$ \\\cline{1-1}
  $4px\mathcal{V}_5=V_4-V_3$ \\\cline{1-1}
  $4(px)^2\mathcal{V}_6=-V_1+V_2+V_3+V_4
 + V_5-V_6$ \\\cline{1-1}
 \end{tabular}
\vspace{0.3cm} \caption{Relations between the calligraphic functions
and nucleon vector DA's.}\label{tab:3}
\end{table}
\begin{table}[h]
\centering
\begin{tabular}{|c|} \hline
  $\mathcal{A}_1=A_1$ \\\cline{1-1}
  $2px\mathcal{A}_2=-A_1+A_2-A_3$ \\\cline{1-1}
   $2\mathcal{A}_3=A_3$ \\\cline{1-1}
  $4px\mathcal{A}_4=-2A_1-A_3-A_4+2A_5$ \\\cline{1-1}
  $4px\mathcal{A}_5=A_3-A_4$ \\\cline{1-1}
  $4(px)^2\mathcal{A}_6=A_1-A_2+A_3+A_4-A_5+A_6$ \\\cline{1-1}
 \end{tabular}
\vspace{0.3cm} \caption{Relations between the calligraphic functions
and nucleon axial vector DA's.}\label{tab:4}
\end{table}
\begin{table}[h]
\centering
\begin{tabular}{|c|} \hline
  $\mathcal{T}_1=T_1$ \\\cline{1-1}
  $2px\mathcal{T}_2=T_1+T_2-2T_3$ \\\cline{1-1}
   $2\mathcal{T}_3=T_7$ \\\cline{1-1}
  $2px\mathcal{T}_4=T_1-T_2-2T_7$ \\\cline{1-1}
  $2px\mathcal{T}_5=-T_1+T_5+2T_8$ \\\cline{1-1}
  $4(px)^2\mathcal{T}_6=2T_2-2T_3-2T_4+2T_5+2T_7+2T_8$ \\\cline{1-1}
  $4px \mathcal{T}_7=T_7-T_8$\\\cline{1-1}
  $4(px)^2\mathcal{T}_8=-T_1+T_2 +T_5-T_6+2T_7+2T_8$\\\cline{1-1}
 \end{tabular}
\vspace{0.3cm} \caption{Relations between the calligraphic functions
and nucleon tensor DA's.}\label{tab:5}
\end{table}

  Each distribution amplitude $F(a_ipx)$=  $S_i$,
$P_i$, $V_i$, $A_i$, $T_i$ can be expressed as:
\begin{equation}\label{dependent1}
F(a_ipx)=\int dx_1dx_2dx_3\delta(x_1+x_2+x_3-1) e^{ip
x\Sigma_ix_ia_i}F(x_i)\; .
\end{equation}
where, $x_{i}$ with $i=1,~2$ and $3$ are longitudinal momentum
fractions carried by the participating quarks. Using the nucleon
wave functions, which their explicit expressions  are calculated in
\cite{Lenz}  and the expression for the heavy quark propagator, and
after performing the Fourier transformation, the final expressions
of the correlation functions for both vertexes are found in terms of
the  nucleon DA's in QCD or theoretical side.  For simplicity, we
present the explicit expressions of the nucleon DA's in the
Appendix.

The next step is to  calculate the phenomenological or physical
sides of the correlation functions. Saturating the correlation
functions with
 a complete set of the initial  state, isolating the ground state and performing the
integral over x, we get:
\begin{equation}\label{phys1}
\Pi_{\mu}^{I}(p,q)=\sum_{s}\frac{\langle N(p)\mid
J^{tr,I}_{\mu}(0)\mid \Sigma_{Q}(p+q,s)\rangle\langle
\Sigma_{Q}(p+q,s)\mid \bar J^{\Sigma_{Q}}(0)\mid
0\rangle}{m_{\Sigma_{Q}}^{2}-(p+q)^{2}}+...,
\end{equation}
\begin{equation}\label{phys1111}
\Pi_{\mu}^{II}(p,q)=\sum_{s}\frac{\langle N(p)\mid
J^{tr,II}_{\mu}(0)\mid \Sigma_{Q}(p+q,s)\rangle\langle
\Sigma_{Q}(p+q,s)\mid \bar J^{\Sigma_{Q}}(0)\mid
0\rangle}{m_{\Sigma_{Q}}^{2}-(p+q)^{2}}+...,
\end{equation}
where, the ... denotes the contribution of the higher states and
continuum. The baryonic to the vacuum matrix element of the
interpolating current, i.e.,  $\langle\Sigma_{Q}(p+q,s)\mid \bar
J^{\Sigma_{Q}}(0)\mid 0\rangle$ can   be parameterized in terms of
the residue,   $\lambda_{\Sigma_{Q}}$ as:
\begin{equation}\label{matrixel2}
\langle\Sigma_{Q}(p+q,s)\mid \bar J^{\Sigma_{Q}}(0)\mid
0\rangle=\lambda_{\Sigma_{Q}} \bar u_{\Sigma_{Q}}(p+q,s).
\end{equation}
 To proceed, we also need to know the transition  matrix elements, $\langle N(p)\mid
J_{\mu}^{tr,I}\mid \Sigma_{Q}(p+q,s)\rangle$ and  $\langle N(p)\mid
J_{\mu}^{tr,II}\mid \Sigma_{Q}(p+q,s)\rangle$. In full theory, they
are parameterized in terms of 12 transition form factors, $f_{i}$,
  $g_{i}$, $f^T_{i}$ and $g^T_{i}$ with $i=1\rightarrow3$ by
the following way:

\begin{eqnarray}\label{matrixel1a}
\langle N(p)\mid J_{\mu}^{tr,I}(x)\mid \Sigma_{Q}(p+q)\rangle&=&\bar
N(p)\left[\gamma_{\mu}f_{1}(Q^{2})+{i}\sigma_{\mu\nu}q^{\nu}f_{2}(Q^{2})+
q^{\mu}f_{3}(Q^{2})-\gamma_{\mu}\gamma_5
g_{1}(Q^{2})-{i}\sigma_{\mu\nu}\gamma_5q^{\nu}g_{2}(Q^{2})\right.\nonumber\\
&-& \left. q^{\mu}\gamma_5 g_{3}(Q^{2})
\vphantom{\int_0^{x_2}}\right] u_{\Sigma_{Q}}(p+q),\nonumber\\
\end{eqnarray}
and
\begin{eqnarray}\label{matrixel1b}
\langle N(p)\mid J_{\mu}^{tr,II}(x)\mid
\Sigma_{Q}(p+q)\rangle&=&\bar
N(p)\left[\gamma_{\mu}f_{1}^{T}(Q^{2})+{i}\sigma_{\mu\nu}q^{\nu}f_{2}^{T}(Q^{2})+
q^{\mu}f_{3}^{T}(Q^{2})+\gamma_{\mu}\gamma_5
g_{1}^{T}(Q^{2})+{i}\sigma_{\mu\nu}\gamma_5q^{\nu}g_{2}^{T}(Q^{2})\right.\nonumber\\
&+& \left. q^{\mu}\gamma_5 g_{3}^{T}(Q^{2})
\vphantom{\int_0^{x_2}}\right] u_{\Sigma_{Q}}(p+q),\nonumber\\
\end{eqnarray}
 where $Q^{2}=-q^{2}$. Here, $N(p)$ and $u_{\Sigma_{Q}}(p+q)$ are
the spinors of nucleon and $\Sigma_{Q}$, respectively.  Using Eqs.
(\ref{phys1}), (\ref{phys1111}), (\ref{matrixel2}),
(\ref{matrixel1a}) and (\ref{matrixel1b}) and performing summation
over spins of the $\Sigma_{Q}$ baryon using
\begin{equation}\label{spinor}
\sum_{s}u_{\Sigma_{Q}}(p+q,s)\overline{u}_{\Sigma_{Q}}(p+q,s)=\not\!p+\not\!q+m_{\Sigma_{Q}},
\end{equation}
 we obtain the following expressions
\begin{eqnarray}\label{phys2}
\Pi_{\mu}^{I}(p,q)&=&
\frac{\lambda_{\Sigma_{Q}}}{m_{\Sigma_{Q}}^{2}-(p+q)^{2}}\bar
N(p)\left[\gamma_{\mu}f_{1}(Q^{2})+{i}\sigma_{\mu\nu}q^{\nu}f_{2}(Q^{2})+
q^{\mu}f_{3}(Q^{2})-\gamma_{\mu}\gamma_5
g_{1}(Q^{2})-{i}\sigma_{\mu\nu}\gamma_5q^{\nu}g_{2}(Q^{2})\right.\nonumber\\
&-& \left. q^{\mu}\gamma_5 g_{3}(Q^{2})
\vphantom{\int_0^{x_2}}\right] (\not\!p+\not\!q+m_{\Sigma_{Q}})
 +
\cdots
\end{eqnarray}
and \begin{eqnarray}\label{phys22} \Pi_{\mu}^{II}(p,q)&=&
\frac{\lambda_{\Sigma_{Q}}}{m_{\Sigma_{Q}}^{2}-(p+q)^{2}}\bar
N(p)\left[\gamma_{\mu}f_{1}^{T}(Q^{2})+{i}\sigma_{\mu\nu}q^{\nu}f_{2}^{T}(Q^{2})+
q^{\mu}f_{3}^{T}(Q^{2})+\gamma_{\mu}\gamma_5
g_{1}^{T}(Q^{2})+{i}\sigma_{\mu\nu}\gamma_5q^{\nu}g_{2}^{T}(Q^{2})\right.\nonumber\\
&+& \left. q^{\mu}\gamma_5 g_{3}^{T}(Q^{2})
\vphantom{\int_0^{x_2}}\right] (\not\!p+\not\!q+m_{\Sigma_{Q}}) +
\cdots
\end{eqnarray}
 Using the relation
\begin{eqnarray}\label{sigma}
\bar{N}\sigma_{\mu\nu}q^{\nu}u_{\Sigma_{Q}}&=i&
\bar{N}[(m_N+m_{\Sigma_{Q}})\gamma_{\mu}-(2p+q)_\mu]u_{\Sigma_{Q}},
\end{eqnarray}
in Eqs. (\ref{phys2}) and Eqs. (\ref{phys22}), we attain the
 final expressions  for the physical side of the correlation
functions:
\begin{eqnarray}\label{sigmaaftera}
\Pi_{\lambda}^{I}(p,q)&=&
\frac{\lambda_{\Sigma_{Q}}}{m_{\Sigma_{Q}}^{2}-(p+q)^{2}}\bar
N(p)\left[\vphantom{\int_0^{x_2}}2f_{1}(Q^{2})p_\mu+\left\{\vphantom{\int_0^{x_2}}-f_1(Q^2)(m_N-m_{\Sigma_{Q}})
+f_2(Q^2)(m_N^2-m_{\Sigma_{Q}}^2)\right\}\gamma_\mu\right.\nonumber \\
&&+\left\{\vphantom{\int_0^{x_2}}f_1(Q^2)-f_2(Q^2)(m_N+m_{\Sigma_{Q}})\right\}\gamma_\mu\not\!q+
2f_{2}(Q^{2})p_\mu\not\!q
+\left\{\vphantom{\int_0^{x_2}}f_2(Q^2)+f_3(Q^2)\right\}(m_N+m_{\Sigma_{Q}})q_\mu\nonumber\\&+&\left\{\vphantom{\int_0^{x_2}}f_2(Q^2)
+f_3(Q^2)\right\}q_\mu\not\!q+ 2g_1(Q^2)p_{\mu}\gamma_5
-\left\{\vphantom{\int_0^{x_2}}g_1(Q^2)(m_N+m_{\Sigma_{Q}})\right.-\left.g_2(Q^2)(m_N^2-m^2_{\Sigma_{Q}})\vphantom{\int_0^{x_2}}\right\}
\gamma_\mu\gamma_5+\nonumber\\&&
\left\{\vphantom{\int_0^{x_2}}g_{1}(Q^{2})-g_2(Q^2)(m_N-m_{\Sigma_{Q}})\right\}\gamma_\mu\not\!q\gamma_5+2g_2(Q^2)p_\mu\not\!q\gamma_5
+\left\{\vphantom{\int_0^{x_2}}g_2(Q^2)+g_3(Q^2)\right\}(m_N-m_{\Sigma_{Q}})q_\mu\gamma_5\nonumber\\&&
+\left\{g_2(Q^2)+g_3(Q^2)\vphantom{\int_0^{x_2}}\right\}q_\mu\not\!q\gamma_5\left.\vphantom{\int_0^{x_2}}\right]+
\cdots
\end{eqnarray}
and
\begin{eqnarray}\label{sigmaafterb}
\Pi_{\lambda}^{II}(p,q)&=&
\frac{\lambda_{\Sigma_{Q}}}{m_{\Sigma_{Q}}^{2}-(p+q)^{2}}\bar N(p)
\left[\vphantom{\int_0^{x_2}}2f_{1}^{T}(Q^{2})p_\mu+\left\{\vphantom{\int_0^{x_2}}-f_1^{T}(Q^2)(m_N-m_{\Sigma_{Q}})
+f_2^{T}(Q^2)(m_N^2-m_{\Sigma_{Q}}^2)\right\}\gamma_\mu\right.\nonumber \\
&&+\left\{\vphantom{\int_0^{x_2}}f_1^{T}(Q^2)-f_2^{T}(Q^2)(m_N+m_{\Sigma_{Q}})\right\}\gamma_\mu\not\!q+
2f_{2}^{T}(Q^{2})p_\mu\not\!q
+\left\{\vphantom{\int_0^{x_2}}f_2^{T}(Q^2)+f_3^{T}(Q^2)\right\}(m_N+m_{\Sigma_{Q}})q_\mu\nonumber\\&+&\left\{\vphantom{\int_0^{x_2}}f_2^{T}(Q^2)
+f_3^{T}(Q^2)\right\}q_\mu\not\!q- 2g_1^{T}(Q^2)p_{\mu}\gamma_5
+\left\{\vphantom{\int_0^{x_2}}g_1^{T}(Q^2)(m_N+m_{\Sigma_{Q}})\right.-\left.g_2^{T}(Q^2)(m_N^2-m^2_{\Sigma_{Q}})\vphantom{\int_0^{x_2}}\right\}
\gamma_\mu\gamma_5-\nonumber\\&&
\left\{\vphantom{\int_0^{x_2}}g_{1}^{T}(Q^{2})-g_2^{T}(Q^2)(m_N-m_{\Sigma_{Q}})\right\}\gamma_\mu\not\!q\gamma_5-2g_2^{T}(Q^2)p_\mu\not\!q\gamma_5
-\left\{\vphantom{\int_0^{x_2}}g_2^{T}(Q^2)+g_3^{T}(Q^2)\right\}(m_N-m_{\Sigma_{Q}})q_\mu\gamma_5\nonumber\\&&
-\left\{g_2^{T}(Q^2)+g_3^{T}(Q^2)\vphantom{\int_0^{x_2}}\right\}q_\mu\not\!q\gamma_5\left.\vphantom{\int_0^{x_2}}\right]+
\cdots
\end{eqnarray}

In order to calculate the form factors or their combinations,
$f_{1}$, $f_{2}$, $f_{2}+f_{3}$, $g_{1}$, $g_{2}$ and
$g_{2}+g_{3}$, we will choose the independent structures $p_{\mu}$,
$p_{\mu}\!\!\not\!q$, $q_{\mu}\!\!\not\!q$, $p_{\mu}\gamma_5$,
$p_{\mu}\!\!\not\!q\gamma_5$, and $q_{\mu}\!\!\not\!q\gamma_5$ from
 Eq. (\ref{sigmaaftera}), respectively. The same structures are chosen to
 calculate the form factors or their combinations labeled by T in the second
 correlation function in Eq. (\ref{sigmaafterb}).

Having computed both sides of the correlation functions, it is time
to obtain the sum rules for the related form factors. Equating the
coefficients of the  corresponding structures from both sides of the
correlation functions through the dispersion relations and applying
Borel transformation with respect to $(p+q)^2$ to suppress the
contribution of the higher states and continuum, one can obtain sum
rules for the  form factors $f_{1}$, $f_{2}$, $f_{3}$, $g_{1}$,
$g_{2}$,  $g_{3}$, $f^T_{1}$, $f^T_{2}$, $f^T_{3}$, $g^T_{1}$,
$g^T_{2}$ and $g^T_{3}$.  In heavy quark effective theory (HQET), where $m_Q\rightarrow \infty$,
the number of independent form factors  is reduced to two, namely,
$F_1$ and $F_2$. In this limit, the transition matrix element can be
parameterized in terms of these two form factors in the following
way \cite{Mannel,alievozpineci}:
\begin{eqnarray}\label{matrixel1111}
\langle N(p)\mid \bar d\Gamma b\mid \Sigma_Q(p+q)\rangle&=&\bar
N(p)[F_1(Q^2)+\not\!vF_2(Q^2)]\Gamma u_{\Sigma_Q}(p+q),\nonumber\\
\end{eqnarray}
where, $\Gamma$ is any Dirac matrices and
$\not\!v=\frac{\not\!p+\not\!q}{m_{\Sigma_{Q}}}$.  Here we should mention that the above relation is exact for $\Lambda$-like baryons, where the light degrees of freedom are spinless. 
For the $\Sigma$ like baryons this relation cannot hold exactly and has to be replaced by a more complicated relation. In the present work, we will use the above approximate relation for the considered transitions.
Comparing this
matrix element and  our definitions of the form factors in Eqs.
(\ref{matrixel1a}) and (\ref{matrixel1b}), we get the following
relations among the form factors in HQET limit (see also
\cite{Chen,ozpineci})
\begin{eqnarray}\label{matrixel22222}
 f_{1}&=&g_{1}=f_{2}^{T} = g_{2}^{T}=F_1+ \frac{m_N}{m_{\Lambda_b}}F_2\nonumber\\
f_2 &=& g_2 = f_3 = g_3=\frac{F_2}{m_{\Sigma_Q}}\nonumber\\
f_1^{T} &=& g_1^{T} =\frac{F_2}{m_{\Sigma_Q}}q^2\nonumber\\
f_3^{T} &=&-\frac{F_2}{m_{\Sigma_Q} }(m_{\Sigma_Q}-m_{N})\nonumber\\
g_3^{T} &=&\frac{F_2}{m_{\Sigma_Q} }(m_{\Sigma_Q}+m_{N})
\end{eqnarray}
Looking at the above relations, we see that it is possible to write
all form factors in terms of $f_1$ and $f_2$, so  we will present
the explicit expressions for these two form factors in the Appendix and give
extrapolation of the other  form factors in finite mass as well as
HQET in terms of $q^2$ in the numerical analysis section.

The  expressions of the sum rules for  form factors show that we
need to know also the residue $\lambda_{\Sigma_{Q}}$. This residue
is determined in \cite{Ozpineci1}:
\begin{eqnarray}\label{residu2}
-\lambda_{\Sigma_{Q}}^{2}e^{-m_{\Sigma_{Q}}^{2}/M_B^{2}}&=&\int_{m_{Q}^{2}}^{s_{0}}e^{\frac{-s}{M_B^{2}}}\rho(s)ds+e^{\frac{-m_Q^2}{M_B^{2}}}\Gamma,
\end{eqnarray}
where,
\begin{eqnarray}\label{residurho1}
\rho(s)&=&(<\overline{d}d>+<\overline{u}u>)\frac{(\beta^{2}-1)}{64
\pi^{2}}\Bigg\{\frac{m_{0}^{2}}{4 m_{Q}}
(6\psi_{00}-13\psi_{02}-6\psi_{11})+3m_{Q}(2\psi_{10}-\psi_{11}-\psi_{12}+2\psi_{21})\Bigg\}
\nonumber\\&+&\frac{ m_{Q}^{4}}{2048 \pi^{4}}
[5+\beta(2+5\beta)][12\psi_{10}-6\psi_{20}+2\psi_{30}-4\psi_{41}+\psi_{42}
-12 ln(\frac{s}{m_{Q}^{2}})],\nonumber\\
\end{eqnarray}
and
\begin{eqnarray}\label{lamgamma1}
\Gamma&=&\frac{
(\beta-1)^{2}}{24}<\overline{d}d><\overline{u}u>\left[\vphantom{\int_0^{x_2}}\right.\frac{m_{Q}^{2}m_{0}^{2}}{2
M_B^{4}} +\frac{m_{0}^{2}}{4 M_B^{2}}-1\Bigg].
\end{eqnarray}
Here,  $\psi_{nm}=\frac{(s-m_Q^2)^n}{s^m(m_Q^2)^{n-m}}$ are some
dimensionless functions.

\section{Numerical results}
This section deals with the numerical analysis of the form factors
  as well as the total decay rate
of  the loop level $\Sigma_{b}\longrightarrow n \ell^+\ell^-$ and
$\Sigma_{c}\longrightarrow p \ell^+\ell^-$ transitions in both full
theory and HQET limit. In obtaining numerical values, we use the
following  inputs for masses and quark condensates: $\uu(1~GeV) =
\dd(1~GeV)= -(0.243)^3~GeV^3$, $m_n = 0.939~GeV$, $m_p = 0.938~GeV$,
$m_b = 4.7~GeV$, $m_c = 1.23~GeV$, $m_{\Sigma_{b}} = 5.805~GeV$,
$m_{\Sigma_{c}} = 2.4529~GeV$ and $m_0^2(1~GeV) = (0.8\pm0.2)~GeV^2$
\cite{Belyaev}. From the sum rules expressions for the form factors,
it is clear that  the nucleon DA's (see Appendix) are the main
input parameters.  These DA's contain eight independent parameters,
namely, $f_{N},~\lambda_{1},
~\lambda_{2},~V_{1}^{d},~A_{1}^{u},~f_{1}^{d},~f_{1}^{u}$ and
$f_{2}^{d}$. All of these parameters have been calculated in the
framework of the light cone QCD sum rules \cite{Lenz} and  most of
them  are now available in  lattice QCD
\cite{Gockeler1,Gockeler2,QCDSF} (see Table \ref{kazem}). Here, we should stress that in \cite{Lenz} those parameters are obtained both as QCD sum rules  and assymptotic sets, but
to improve the agreement with experimental data on nucleon form factors, a set of parameters is obtained  using a simple
model in which the deviation from the asymptotic
DAs is taken to be 1/3 of that suggested by
the QCD sum rule estimates (see \cite{Lenz}). We will use this set of parameters in this paper and refer it as set1 (see Table \ref{kazem}). In the following, we also will denote the lattice QCD input parameters by
 set2.

\begin{table}[h]
\centering
\begin{tabular}{|c||c|c|} \hline
& set1 \cite{Lenz} & set2 or Lattice QCD
\cite{Gockeler1,Gockeler2,QCDSF}
\\\cline{1-3} \hline\hline
$f_{N}$ & $(5.0\pm0.5)\times10^{-3}~GeV^{2}$ &
$(3.234\pm0.063\pm0.086)\times10^{-3}~GeV^{2}$
\\\cline{1-3} $\lambda_{1}$ &$-(2.7\pm0.9)\times10^{-2}~GeV^{2}$ & $(-3.557\pm0.065\pm0.136)\times10^{-2}~GeV^{2}$ \\\cline{1-3}
 $\lambda_{2}$
&$(5.4\pm1.9)\times10^{-2}~GeV^{2}$&
$(7.002\pm0.128\pm0.268)\times10^{-2}~GeV^{2}$\\\cline{1-3}
$V_{1}^{d}$ &$0.30$& $0.3015\pm0.0032\pm0.0106$
\\\cline{1-3}
$A_{1}^{u}$ &$0.13$& $0.1013\pm0.0081\pm0.0298$\\\cline{1-3}
$f_{1}^{d}$ &$0.33$& $-$\\\cline{1-3} $f_{1}^{u}$
&$0.09$& $-$\\\cline{1-3} $f_{2}^{d}$ &$0.25$&
$-$\\\cline{1-3}
\end{tabular}
\vspace{0.8cm} \caption{The values of the 8 independent parameters
entering  the nucleon DA's. The first errors in lattice values are
statistical and the second errors correspond to  the uncertainty due
to the Chiral extrapolation and renormalization. For last tree
parameters, the values are not available in lattice and we will use
the set1 values for both sets of data.} \label{kazem}
\end{table}

The explicit expressions for the form factors also show their
dependency to three auxiliary mathematical  objects, namely,
continuum threshold $s_0$, Borel mass parameter $M_B^2$ and general
parameter $\beta$ entering to the most general form of the
interpolating current of the initial state. The form factors as
physical quantities should be independent of these parameters, hence
we need to look for working regions for them. The working region for
Borel mass squared is determined as follows:  the upper limit of
$M_B^2$ is chosen demanding that  the series of the light cone
expansion with increasing twist should be convergent. The lower
limit  is determined from condition that the higher states and
continuum contributions constitute a small fraction of total
dispersion integral. Both conditions are satisfied in the regions
$15 ~GeV^{2}\leq M_{B}^{2}\leq 30~ GeV^{2}$ and    $4 ~GeV^{2}\leq
M_{B}^{2}\leq 12~ GeV^{2}$ for  bottom and charm cases,
respectively. The value of the continuum threshold $s_{0}$ is not
completely arbitrary and it is correlated to the first exited state
with quantum numbers of the initial particle interpolating current.
Our numerical calculations show that the form factors weakly depend
on the continuum threshold in the interval,
$(m_{\Sigma_Q}+0.5)^2\leq s_0\leq (m_{\Sigma_Q}+0.7)^2$. To obtain
the working region for $\beta$ at which the form factors are
practically independent of it, we look for the variation of the form
factors with respect to  $ cos\theta$ in the interval $-1\leq
cos\theta\leq1$ which is equivalent  to the $-\infty\leq
\beta\leq\infty$, where $\beta=tan\theta$. As a result, the interval
$-0.5\leq cos\theta\leq0.6$ is obtained for  $\beta$  for both charm
and bottom cases. In this interval, the dependency on this parameter is weak.

  The next step is to discuss the behaviour of the form factors in terms of the $q^2$.  The sum rules predictions for the form factors are not reliable
in the whole physical region. To be able to extend the results for the form
factors to the whole physical region, we look for a parametrization of the form factors such that in the reliable region which is approximately $1~ GeV$ below the
perturbative cut,  the original form factors and their fit parametrization coincide each other.
Our numerical results lead to the following
extrapolation for the form factors in terms of $q^2$:

\begin{equation}\label{17au}
 f_{i}(q^2)[g_{i}(q^2)]=\frac{a}{(1-\frac{q^2}{m^2_{fit}})}+\frac{b}{(1-\frac{q^2}{m^2_{fit}})^2},
\end{equation}
 where the fit parameters
$a,~b$ and $m_{fit}$ in full theory  and  HQET limit are given in
Tables \ref{tab:13}, \ref{tab:14}, \ref{tab:31} and \ref{tab:41}
using two sets for the
independent  parameters.  These Tables, show  poles of the form
factors  outside the allowed physical region.  Therefore, the form
factors are analytic in the full physical interval. In principle, we can use fit parametrization either with single pole or double poles. However, when we combine them 
the accuracy of the fitting becomes very high, specially  when the pole is the same for two parts.  We could start 
 from $f_i(q^2)=\frac{a}{1-q^2/m_1^2}+\frac{b}{1-q^2/m_{fit}^2}$, however for all form factors $m_{fit}$ gets too close to $m_1$, so  the fit becomes numerically
 unstable. In such a case,  it is appropriate to expand the above relation to first order
in $m_{fit}-m_1$, which gives the Eq. (\ref{17au}) used to extrapolate the factors over the whole range of $q^2$. For the same situation in 
$B \longrightarrow D$ mesonic transition see for instance \cite{Becirevic,Ballb}. The values of
form factors at $q^2=0$  are
presented in Tables \ref{tab:15} and \ref{tab:16} in both full
theory and HQET for bottom and charm cases, respectively. In extraction of the values of form factors at $q^2=0$, the mean values of the form factors obtained from the quoted ranges for the auxiliary parameters
have been considered.  When we look at these Tables, we see that although the values for the eight independent input parameters for two sets are close to each other but the results
for the central values of some form factors differ in two sets, considerably. The numerical results show that the result of sum rules are very sensitive to these parameters
 specially $f_N$, $\lambda_2$ and $A_1^u$. Within the errors, the quoted values become close to each other for both sets. The
numerical analysis depicts also that all form factors approximately satisfy
 the HQET limit relations in Eq. (\ref{matrixel22222}) within the errors for both
sets of input parameters and bottom case, $Q=b$ at $q^2=0$. However for the charm case, $Q=c$ although some of the relations are satisfied but most of them  are violated at $q^2=0$. This  is an  expected result
since the $m_c\rightarrow \infty$ limit is not as reasonable as the $m_b\rightarrow \infty$.  
\begin{table}[h]
\renewcommand{\arraystretch}{1.5}
\addtolength{\arraycolsep}{3pt}

$$
\begin{array}{|c|c|c|c|c|c|c|}
\hline \hline
     & \multicolumn{3}{c|}{\mbox{set1}} & \multicolumn{3}{c|}{\mbox{set2 }} \\
\hline
     & \mbox{a} & \mbox{b} & \mbox{$m_{fit}$}    & \mbox{a} & \mbox{b} & \mbox{$m_{fit}$}   \\
\hline
 f_1 &  -0.16   &  0.29    & 5.70                &  0.027    &  0.044   & 5.02               \\
 f_2 &  0.008    & -0.02   & 5.96                &   0.018  & -0.024  & 7.96               \\
 f_3 &  0.011    & -0.024   & 6.34                &  -0.003   &  -0.003   & 6.45               \\
 g_1 & -0.21    &  0.33    & 5.73                & -0.13   &  0.20   & 5.29               \\
 g_2 &  0.008     & -0.02    & 5.92                &  -0.01    & 0.003   & 5.72               \\
 g_3 &  0.005     & -0.02    & 5.87                &  0.014    & -0.023    & 7.83               \\
 f_1^{T} &  -0.06      &  0.023    & 5.22                &  -0.029   & -0.018   & 5.12               \\
 f_2^{T} &  -0.16      &  0.29     & 6.47                &  0.069    &  -0.017    & 5.69               \\
 f_3^{T} &  -0.18      &  0.25     & 8.81                &  0.084    &  -0.023    & 5.13               \\
 g_1^{T} &  -0.15      &  0.14     & 5.02                &  -0.01    & -0.028   & 5.11               \\
 g_2^{T} &  -0.20     &  0.31    & 5.24                &  0.026    &  0.04   & 4.72               \\
 g_3^{T} &  0.17       &  -0.25    & 5.76               &   0.11    & -0.18    & 5.33               \\
\hline \hline
\end{array}
$$
\caption{Parameters appearing in  the fit function of the  form
factors in full theory for $\Sigma_{b}\rightarrow n
\ell^{+}\ell^{-}$.} \label{tab:13}
\renewcommand{\arraystretch}{1}
\addtolength{\arraycolsep}{-1.0pt}
\end{table}

\begin{table}[h]
\renewcommand{\arraystretch}{1.5}
\addtolength{\arraycolsep}{3pt}

$$
\begin{array}{|c|c|c|c|c|c|c|}                                  \hline \hline
 & \multicolumn{3}{c|}{\mbox{set1}} & \multicolumn{3}{c|}{\mbox{set2 }}                             \\
 \hline
         & \mbox{a} & \mbox{b} & \mbox{$m_{fit}$}    & \mbox{a} & \mbox{b} & \mbox{$m_{fit}$}                      \\
\hline
 f_{1}   &  0.08    &  0.097     & 1.53                & -0.12    &  0.18    & 1.52                                  \\
 f_{2}   & -0.009   & -0.056    & 1.57                & -0.01   & -0.029   & 1.53                                  \\
 f_{3}   & -0.025   &  0.012    & 1.61                &  0.008   & -0.047    & 1.56                                  \\
 g_{1}   & -0.015   &  0.31     & 1.59                & -0.038   &  0.21    & 1.60                                  \\
 g_{2}   & -0.008   & -0.12     & 1.55                &  0.002   & -0.14    & 1.61                                  \\
 g_{3}   & -0.026    & -0.13     & 1.53                & -0.024   & -0.14    & 1.52                                  \\
 f_{1}^{T}   & -0.23    &  0.19     & 1.52                &  0.09    & -0.097     & 1.58                                  \\
 f_{2}^{T}   &  0.066   &  0.067    & 1.63                &  0.12    &  0.13     & 1.55                                  \\
 f_{3}^{T}   &  0.15    &  0.006     & 1.56                &  0.21    &  0.032     & 1.65                                 \\
 g_{1}^{T}   & -0.45    &  0.29     & 1.59                & -0.17    &  0.09     & 1.62                                  \\
 g_{2}^{T}   &  0.009   &  0.08     & 1.57                & -0.026    &  0.14     & 1.59                                  \\
 g_{3}^{T}   & -0.09    & -0.11     & 1.54                & -0.07    & -0.16     & 1.56                                  \\
\hline \hline
\end{array}
$$
\caption{Parameters appearing in   the fit function of the form
factors in full theory for $\Sigma_{c}\rightarrow p
\ell^{+}\ell^{-}$.} \label{tab:14}
\renewcommand{\arraystretch}{1}
\addtolength{\arraycolsep}{-1.0pt}
\end{table}

\begin{table}[h]
\renewcommand{\arraystretch}{1.5}
\addtolength{\arraycolsep}{3pt}

$$
\begin{array}{|c|c|c|c|c|c|c|}
\hline \hline
     & \multicolumn{3}{c|}{\mbox{set1}} & \multicolumn{3}{c|}{\mbox{set2 }} \\
\hline
     & \mbox{a} & \mbox{b} & \mbox{$m_{fit}$}    & \mbox{a} & \mbox{b} & \mbox{$m_{fit}$}   \\
\hline
  f_1 &  -0.22   &  0.4   & 4.96                & 0.037   &  0.06   & 5.13               \\
 f_2 &    0.009   & -0.024   & 5.13                &  0.021   & -0.028   & 5.28               \\
 f_3 &    0.009   & -0.02  & 5.72                &  -0.003   & -0.002   & 5.67               \\
 g_1 &   -0.029   &  0.19   & 5.32                & -0.18   &  0.28    & 5.57               \\
 g_2 &    0.008   & -0.02  & 5.41                &  -0.01   & 0.003    & 5.35               \\
 g_3 &    0.005   & -0.018   & 4.87                &  0.013   & -0.021    & 5.06               \\
 f_1^{T} &  -0.065      &   0.025    & 5.16                &  -0.03    & 0.019    & 5.25               \\
 f_2^{T} &   -0.24     &  0.43     & 5.04                &   0.104    & -0.026    & 5.13               \\
 f_3^{T} &   -0.19     &  0.27    & 5.13                &   0.091    & -0.025    & 5.54               \\
 g_1^{T} &  -0.14      &  0.13      & 5.11                &  -0.009    &  - 0.011    & 5.14               \\
 g_2^{T} &  -0.03     &  0.17     & 5.58                &  0.04    &  0.06    & 5.47               \\
 g_3^{T} &   0.21     &  -0.27    & 5.16                &   0.1    & -0.17    & 4.96               \\
\hline \hline
\end{array}
$$
\caption{Parameters appearing in the fit function of the form
factors at HQET limit  for $\Sigma_{b}\rightarrow n
\ell^{+}\ell^{-}$.} \label{tab:31}
\renewcommand{\arraystretch}{1}
\addtolength{\arraycolsep}{-1.0pt}
\end{table}
\begin{table}[h]
\renewcommand{\arraystretch}{1.5}
\addtolength{\arraycolsep}{3pt}

$$
\begin{array}{|c|c|c|c|c|c|c|}                                  \hline \hline
 & \multicolumn{3}{c|}{\mbox{set1}} & \multicolumn{3}{c|}{\mbox{set2 }}                             \\
 \hline
         & \mbox{a} & \mbox{b} & \mbox{$m_{fit}$}    & \mbox{a} & \mbox{b} & \mbox{$m_{fit}$}                      \\
\hline
 f_{1}   & 0.02    &  0.13    & 1.64                & -0.17    &  0.25    & 1.55                                  \\
 f_{2}   & -0.011   & -0.077   & 1.76                & -0.014   & -0.04    & 1.51                                  \\
 f_{3}   & -0.034   & 0.017    & 1.73                & 0.011   &  -0.065    & 1.62                                  \\
 g_{1}   & -0.021   &  0.43    & 1.68                & -0.053   &  0.29    & 1.65                                  \\
 g_{2}   & -0.008   & -0.12    & 1.57                & -0.002   & -0.14    & 1.63                                  \\
 g_{3}   & -0.26   & 0.10    & 1.62                & -0.022   & -0.13    & 1.57                                  \\
 f_{1}^{T}   & -0.097    &  0.05     & 1.58                &  0.098    &  -0.1     & 1.65                                  \\
 f_{2}^{T}   & 0.099   &  0.1     & 1.69                & 0.18    &  0.196     & 1.59                                  \\
 f_{3}^{T}   &  0.14   &  0.009    & 1.51                &  0.32    &  -0.048     & 1.71                                  \\
 g_{1}^{T}   & -0.4    &  0.26     & 1.66                & -0.15     & 0.08     & 1.57                                  \\
 g_{2}^{T}   & 0.014   &  0.12     & 1.63                & -0.039    &  0.21     & 1.63                                  \\
 g_{3}^{T}   & -0.08   & -0.1     & 1.58                &  -0.064    & -0.15     & 1.60                                  \\
\hline \hline
\end{array}
$$
\caption{Parameters appearing in the fit function of the form
factors at HQET limit for  $\Sigma_{c}\rightarrow p
\ell^{+}\ell^{-}$.} \label{tab:41}
\renewcommand{\arraystretch}{1}
\addtolength{\arraycolsep}{-1.0pt}
\end{table}

\begin{table}[h]
\renewcommand{\arraystretch}{1.5}
\addtolength{\arraycolsep}{3pt}

$$
\begin{array}{|c|c|c|c|c|}                                  \hline \hline
 & \multicolumn{2}{c|}{\mbox{Full Theory}} & \multicolumn{2}{c|}{\mbox{HQET }}                    \\
 & \multicolumn{1}{c}{\mbox{set1}} & \multicolumn{1}{c|}{\mbox{set2}}
 & \multicolumn{1}{c} {\mbox{set1}}
 & \multicolumn{1}{c|}{\mbox{set2 }}                            \\ \hline
 f_{1}(0)     & 0.14 \pm0.04    &  0.07 \pm0.02    &  0.19 \pm0.05    &   0.10 \pm0.03      \\
 f_{2}(0)     &-0.012 \pm0.003   & -0.006\pm0.002   & -0.014\pm0.004   &  -0.005\pm0.002     \\
 f_{3}(0)     &-0.013\pm0.003   & -0.006\pm0.002  & -0.011 \pm0.002   &  -0.005 \pm0.002       \\
 g_{1}(0)     & 0.12\pm0.03   & 0.07 \pm0.02   &  0.17 \pm0.04   &   0.10 \pm0.03       \\
 g_{2}(0)     &-0.012\pm0.003   & -0.007\pm0.002   & -0.012\pm0.004   &  -0.007 \pm0.002         \\
 g_{3}(0)     &-0.014 \pm0.004   & -0.009 \pm0.003    & -0.013\pm0.004   &  -0.008\pm0.003          \\
 f_{1}^{T}(0)     & -0.03\pm0.01    &  -0.04\pm0.01   &  -0.03\pm0.01    &   -0.010\pm0.003      \\
 f_{2}^{T}(0)     &  0.13\pm0.04   &   0.052\pm0.020   &   0.19\pm0.05   &    0.079\pm0.003     \\
 f_{3}^{T}(0)     &  0.07\pm0.02    &   0.061\pm0.020   &   0.08\pm0.03   &    0.066\pm0.021       \\
 g_{1}^{T}(0)     & -0.012\pm0.003   &  -0.03\pm0.01  &  -0.012\pm0.004   &   -0.020\pm0.006       \\
 g_{2}^{T}(0)     &  0.11\pm0.03   &   0.066\pm0.021  &   0.16\pm0.04   &    0.10\pm0.03         \\
 g_{3}^{T}(0)     & -0.07\pm0.02     &  -0.073\pm0.025   &  -0.07\pm0.02   &   -0.066\pm0.021         \\
 \hline \hline
\end{array}
$$

\caption{The values of the form factors at $q^2=0$  for
$\Sigma_{b}\rightarrow n \ell^{+}\ell^{-}$.} \label{tab:15}
\renewcommand{\arraystretch}{1}
\addtolength{\arraycolsep}{-1.0pt}
\end{table}

\begin{table}[h]
\renewcommand{\arraystretch}{1.5}
\addtolength{\arraycolsep}{3pt}

$$
\begin{array}{|c|c|c|c|c|}                                  \hline \hline
 & \multicolumn{2}{c|}{\mbox{Full Theory}} & \multicolumn{2}{c|}{\mbox{HQET }}                    \\
 & \multicolumn{1}{c}{\mbox{set1}} & \multicolumn{1}{c|}{\mbox{set2}}
 & \multicolumn{1}{c} {\mbox{set1}}
 & \multicolumn{1}{c|}{\mbox{set2 }}                            \\ \hline
 f_{1}(0)     &  0.19 \pm0.05   & 0.05\pm0.02    &  0.16\pm0.05     &  0.069\pm0.022  \\
 f_{2}(0)     & -0.066\pm0.021    & -0.04\pm0.01   & -0.078\pm0.023    & -0.047\pm0.014  \\
 f_{3}(0)     & -0.013\pm0.003   & -0.039\pm0.012   & -0.010\pm0.003   & -0.034\pm0.011      \\
 g_{1}(0)     &  0.30\pm0.09    & 0.17\pm0.06     &  0.40\pm0.12     &  0.24\pm0.06      \\
 g_{2}(0)     & -0.12\pm0.03    & -0.14\pm0.04    & -0.12\pm0.04    & -0.14\pm0.04      \\
 g_{3}(0)     & -0.16\pm0.05    & -0.16\pm0.05    & -0.15\pm0.05     & -0.15\pm0.04       \\
 f_{1}^{T}(0)     & -0.039\pm0.012    &  -0.007\pm0.002 &  -0.042\pm0.013    &   -0.0020\pm0.0007       \\
 f_{2}^{T}(0)     &  0.14\pm0.04    &   0.25\pm0.07    &   0.21\pm0.06      &    0.38\pm0.12      \\
 f_{3}^{T}(0)     &  0.15\pm0.05    &   0.24\pm0.07    &   0.16\pm0.04      &    0.26\pm0.08       \\
 g_{1}^{T}(0)     & -0.16\pm0.05    &  -0.08\pm0.03   &  -0.14\pm0.05      &   -0.07\pm0.02       \\
 g_{2}^{T}(0)     &  0.09\pm0.03    &   0.10\pm0.03   &   0.14\pm0.05      &    0.15\pm0.05         \\
 g_{3}^{T}(0)     & -0.20\pm0.07    &  -0.23\pm0.06   &  -0.18\pm0.05      &   -0.21\pm0.08          \\

 \hline \hline
\end{array}
$$

\caption{The values of the form factors  at $q^2=0$ for
$\Sigma_{c}\rightarrow p \ell^{+}\ell^{-}$.} \label{tab:16}
\renewcommand{\arraystretch}{1}
\addtolength{\arraycolsep}{-1.0pt}
\end{table}

 Our next task is to calculate the total decay rate of the FCNC $\Sigma_{b}\longrightarrow p \ell^+\ell^-$
 and $\Sigma_{c}\longrightarrow n \ell^+\ell^-$ transitions in the
 full allowed
physical region, namely, $ 4m_{l}^2 \leq q^2 \leq (m_{\Sigma_{b,c}}
- m_{p,n})^2$.  To derive the expression for the decay rate, we will
make the following assumptions (see also \cite{Chuan}): the CLEO
predicts the value $R=\frac{F_2}{F_1}=-0.25\pm0.14\pm0.08$ for the
ratio of the form factors of $ \Lambda_c \rightarrow \Lambda ee_\nu$
at HQET limit \cite{CLEO}. This result shows that $|F_2|<|F_1|$ and
considering  Eq. (\ref{matrixel22222}), the form factors $f_1$,
$g_1$, $f_1$, $f^T_2$ and $g^T_2$ are expected to be large comparing
to the other form factors since they are proportional to the $F_1$.
Moreover, it is clear from the considered Hamiltonian as well as the
definition of the transition matrix elements in terms of the form
factors that the form factors labeled by T  are related to the
Wilson coefficient $C_7$ which is about one order of  magnitude
smaller than the other coefficients entered to the Hamiltonian,
i.e., $C_9$ and $C_{10}$, hence their effects expected to be small.
As a result of the above procedure,  the following results for decay
width describing such transitions is obtained \cite{Chuan}:
\begin{equation}
\frac{d\Gamma } {ds}\left( \Sigma _{Q}\rightarrow N l^{+}l^{-}\right)=%
\frac{G_{F}^{2}\alpha _{em}^{2}|V_{Q'Q}~V^*_{Q'q}|^{2}}{384\pi
^{5}}m_{\Sigma
_{Q}}^{5}\sqrt{\phi \left( s\right) }\sqrt{1-\frac{4m_{l}^{2}}{q^{2}}}\bar{f}%
^{2}R_{\Sigma _{Q}}\left( s\right) ,  \label{rate}
\end{equation}
where
\begin{equation}
 R_{\Sigma _{Q}}\left( s\right) =\Gamma _{1}\left(
s\right) +\Gamma _{2}\left( s\right) +\Gamma _{3}\left( s\right)
\end{equation}
and
\begin{eqnarray}
\Gamma _{1}\left( s\right) &=&-6\sqrt{r}s\left[ -2\hat{m}_{Q}\rho \left( 1+2%
\frac{m_{l}^{2}}{q^{2}}\right) {\rm {Re}C_{9}^{eff}C_{7}^{*}}\right.
\left. +\delta \left( \left( 1+2\frac{m_{l}^{2}}{q^{2}}\right)
\left| C_{9}^{eff}\right| ^{2}+\left(
1-6\frac{m_{l}^{2}}{q^{2}}\right) \left|
C_{10}\right| ^{2}\right) \right]  \nonumber \\
&&+\left[ -2r\left( 1+2\frac{m_{l}^{2}}{q^{2}}\right) -4t^{2}\left( 1-\frac{%
m_{l}^{2}}{q^{2}}\right) +3\left( 1+r\right) t\right]  \times \left[
\left( 2\hat{m}_{Q}\rho \right) ^{2}\left| C_{7}^{}\right|
^{2}+\left| C_{9}^{eff}\right| ^{2}+\left| C_{10}\right|
^{2}\right]  \nonumber \\
&&+6\hat{m}_{l}^{2}t\left[ \left( 2\hat{m}_{Q}\rho \right)
^{2}\left| C_{7}^{}\right| ^{2}+\left| C_{9}^{eff}\right|
^{2}-\left| C_{10}\right|
^{2}\right] ,  \label{rate1} \\
\Gamma _{2}\left( s\right) &=&6\sqrt{r}\left( 1-t\right) \left\{ 4\left( 1+2%
\frac{m_{l}^{2}}{q^{2}}\right) \hat{m}_{Q}^{2}\rho \left|
C_{7}^{}\right| ^{2}\right. \left. +\rho s\left[ \left(
1+2\frac{m_{l}^{2}}{q^{2}}\right) \left| C_{9}^{eff}\right|
^{2}+\left( 1-2\frac{m_{l}^{2}}{q^{2}}\right) \left|
C_{10}\right| ^{2}\right] \right\}  \nonumber \\
&&+12\left( 1+2\frac{m_{l}^{2}}{q^{2}}\right) \hat{m}_{Q}\left(
t-r\right) \left( 1+s\rho ^{2}\right) {\rm
{Re}C_{9}^{eff}C_{7}^{*}},  \label{rate2}
\\
\Gamma _{3}\left( s\right) &=&12\left(
1+2\frac{m_{l}^{2}}{q^{2}}\right)
\hat{m}_{Q}\sqrt{r}s\rho {\rm {Re}C_{9}^{eff}C_{7}^{*}}
 -\left[ 2t^{2}\left( 1+2\frac{m_{l}^{2}}{q^{2}}\right) +4r\left( 1-\frac{%
m_{l}^{2}}{q^{2}}\right) -3\left( 1+r\right) t\right]  \nonumber \\
&&\times \left[ \frac{4\hat{m}_{Q}^{2}}{s}\left| C_{7}\right|
^{2}+s\rho ^{2}\left( \left| C_{9}^{eff}\right| ^{2}+\left|
C_{10}\right| ^{2}\right)
\right] -6\hat{m}_{l}^{2}\left( 2r-\left( 1+r\right) t\right) \left[ \left( \frac{2%
\hat{m}_{Q}}{s}\right) ^{2}\left| C_{7}^{{\rm }}\right| ^{2}+\rho
^{2}\left( \left| C_{9}^{{\rm eff}}\right| ^{2}-\left| C_{10}\right|
^{2}\right) \right]\,. \nonumber \\ \label{rate3}
\end{eqnarray}
 Here, $G_F = 1.17 \times 10^{-5}$ GeV$^{-2}$ is the Fermi
coupling constant, $ \bar f=\frac{f_1+g_1}{2}$, $
\rho=m_{\Sigma_Q}\frac{f_2+g_2}{f_1+g_1}$,
$\delta=\frac{f_1-g_1}{f_1+g_1}$, $s=\frac{q^2}{m^2_{\Sigma_Q}}$,
$\hat{m}_{Q}=\frac{m_Q}{m_{\Sigma_Q}}$,
$\hat{m}_{l}=\frac{m_l}{m_{\Sigma_Q}}$,
$r=\frac{m^2_N}{m^2_{\Sigma_Q}}$,
$t=\frac{1}{2m^2_{\Sigma_Q}}[m^2_{\Sigma_Q}+m^2_N-q^2]$ and $m_l$ is
the lepton  mass. For the Wilson coefficients, we use $C_7=-0.313, C_9=4.344, C_{10}=-4.669$ \cite{Burasb}. Here we should mention that    
the Wilson coefficient $C_9^{eff}$ receives long distance contributions from $J/\psi$ family, in addition to short distance contributions. In the present work, we do
not take into account the long distance effects.
 The elements of the CKM matrix
$V_{tb}=(0.77^{+0.18}_{-0.24})$, $V_{td}=(8.1\pm0.6)\times10^{-3}$,
$V_{bc}=(41.2\pm1.1)\times10^{-3}$ and
$V_{bu}=(3.93\pm0.36)~10^{-3}$
 have also been used~\cite{Yao:2006px}.

 Using the formula for the decay rate the  final results as shown in  Table \ref{tab:27} are obtained.
\begin{table}[h] \centering
 \begin{tabular}{|c||c|c|c|c|c|c|} \hline &
$\Sigma_{b}\longrightarrow n e^+e^-$ &$\Sigma_{b}\longrightarrow n
\mu^+\mu^-$ &$\Sigma_{b}\longrightarrow n \tau^+\tau^-$&
$\Sigma_{c}\longrightarrow p e^+e^- $ & $\Sigma_{c}\longrightarrow p
\mu^+\mu^-$
\\\cline{1-6}\hline\hline
 Full (set1)& $(4.26\pm1.27) \times 10^{-20}$ &$(2.08\pm0.70)\times
10^{-20}$& $(1.0\pm0.3) \times 10^{-22}$&
$(5.59\pm1.78)\times 10^{-25}$& $(9.7\pm2.7) \times 10^{-26}$\\
\cline{1-6}  Full (set2)& $(5.4\pm1.6) \times 10^{-21}$ &
$(2.64\pm0.79) \times 10^{-21}$& $(4.01\pm1.25) \times 10^{-23}$&
$(1.35\pm0.35) \times 10^{-25}$& $(2.36\pm0.80) \times
10^{-26}$\\\cline{1-6}HQET(set1)& $(8.20\pm3.04) \times 10^{-20}$
&$(4.25\pm2.07)\times 10^{-20}$& $(6.26\pm2.46) \times 10^{-22}$&
$(7.99\pm3.07)\times 10^{-25}$& $(1.50\pm0.58) \times 10^{-25}$\\
\cline{1-6} HQET(set2) & $(1.10\pm0.33) \times 10^{-20}$ &
$(5.67\pm1.73) \times 10^{-21}$& $(1.16\pm0.46) \times 10^{-22}$&
$(2.50\pm0.81) \times 10^{-25}$& $(4.30\pm1.36) \times
10^{-26}$\\\cline{1-6}\hline\hline
\end{tabular}
\vspace{0.8cm} \caption{Values of the
$\Gamma(\Sigma_{b,c}\longrightarrow n,p~\ell^+\ell^-$) in  GeV  for
different leptons and two sets of input parameters.} \label{tab:27}
\end{table}
From this table, we see that: a) The value of the decay rate
decreases by increasing in the lepton mass. This is reasonable since
the phase space in for example $\tau$ case is smaller than that of
the electron and $\mu$ cases. b) The order of magnitude on
decay rate of bottom case shows the possibility of the experimental
studies on the $\Sigma_{b}\longrightarrow n~\ell^+\ell^-$
transition, specially the $\mu$ case, at large hadron collider (LHC)
in the near future. The lifetime of the $\Sigma_b$ is not exactly
known yet but if we consider its lifetime approximately the same
order of the b-baryon admixture $(\Lambda_b, \Xi_b, \Sigma_b,
\Omega_b)$ lifetime, which is
$\tau=(1.319^{+0.039}_{0.038})\times10^{-12}~s$ \cite{Yao:2006px},
the branching fraction is obtained in $10^{-7}$ order. Any
measurements in this respect and  comparison of the results with the
predictions of the present work can give essential information about
the nature of the $\Sigma_Q$ baryon, nucleon distribution amplitudes
and search for the new physics beyond the standard model.

\newpage
\section{Acknowledgment}
 K. A.   thanks  TUBITAK, Turkish Scientific and Technological
Research Council, for their partial financial support provided under
the project 108T502. We also thank T. M. Aliev and A. Ozpineci for
their useful discussions.

 \newpage

\section*{Appendix }
In this Appendix,  the explicit expressions for the form factors $f_1$ and $f_2$ for b case as well as the nucleon DA's are
given:
\begin{eqnarray}\label{f_{1}}
&&f_{1}(Q^{2})= \frac{1}{\sqrt{2}\lambda_{\Sigma_b}}
e^{m_{\Sigma_b}^{2}/M_{B}^{2}}\left(\vphantom{\int_0^{x_2}}\int_{t_{0}}^{1}dx_{2}\int_{0}^{1-x_{2}}dx_{1}
e^{-s(x_{2},Q^{2})/M_{B}^{2}}\frac{1}{2\sqrt{2}}\left[\vphantom{\int_0^{x_2}}m_b\left\{\vphantom{\int_0^{x_2}}(1+3\beta){\cal
H}_{19}(x_i)\right.\right.\right. -2(-1+\beta){\cal
H}_{17}(x_i)\nonumber\\&&\left.-(3+\beta){\cal
H}_{5}(x_i)\left.\vphantom{\int_0^{x_2}}\right\}-m_Nx_2\left\{\vphantom{\int_0^{x_2}}{\cal
H}_{1_2,-11,-13,19_2,-5,7}(x_i)+\beta{\cal
H}_{11,13,-17_2,-19_8,3_2,5,-7}(x_i)\vphantom{\int_0^{x_2}}\right\}\vphantom{\int_0^{x_2}}\right]\nonumber\\&&
+\int_{t_0}^1dx_2\int_0^{1-x_2}dx_1\int_{t_0}^{x_2}dt_1e^{-s(t_1,Q^2)/M_{B}^{2}
}\left[\vphantom{\int_0^{x_2}}-\frac{m_N^4
m_b}{M_B^4t_1^3\sqrt{2}}(-1+\beta)x_2{\cal
H}_{22}(x_i)\right.\nonumber\\&&\left.-\frac{m_N^2}{M_B^4t_1^2
2\sqrt{2}}\left\{\vphantom{\int_0^{x_2}}m_N^3x_2\Big[(-1+\beta){\cal
H}_{-10,16}(x_i)+2\beta{\cal
H}_{24}(x_i)\Big]+\Big[4m_N^3x_2+m_b\{Q^2+s(t_1,Q^2)\}(-1+\beta)x_2\right.\right.\nonumber\\&&\left.\left.-m_N^2m_b(-1+\beta)(2+3x_2)\Big]{\cal
H}_{22}(x_i)\vphantom{\int_0^{x_2}}\right\} +\frac{m_N^2}{M_B^4 t_1
2\sqrt{2}}
\left\{\vphantom{\int_0^{x_2}}m_Nx_2\Big[Q^2+s(t_1,Q^2){\cal
H}_{16,-22_4}(x_i)+(-1+\beta){\cal H}_{10}(x_i)
\right.\right.\nonumber\\&&\left.-\beta{\cal H}_{16,24_2}(x_i)\Big]
+m_b(-1+\beta)\Big[Q^2(1+3x_2)+s(t_1,Q^2)(1+x_2)\Big]{\cal
H}_{22}(x_i)+m_N^2m_b\Big[x_2(1+3\beta){\cal
H}_{16}(x_i)\right.\nonumber\\&& \left.+2(-1+\beta){\cal
H}_{24}(x_i) +(3+\beta){\cal H}_{10}(x_i)-(-1+\beta)(3+x_2){\cal
H}_{22}(x_i)\Big] -m_N^3\Big[(-1+\beta)(1+x_2){\cal
H}_{10,-16}(x_i)\right.\nonumber\\&&
-\left.\left.2\{\beta(1+x_2){\cal H}_{24}(x_i)+(2+4x_2){\cal
H}_{22}(x_i)\}\Big]\vphantom{\int_0^{x_2}}\right\}
+\frac{m_N^2}{M_B^42\sqrt{2}}\left\{\vphantom{\int_0^{x_2}}-3m_b
Q^2(-1+\beta){\cal
H}_{22}(x_i)\right.\right.\nonumber\\&&\left.\left.+m_N^2m_b\Big[(-1+\beta){\cal
H}_{22,-24_2}(x_i)-(1+3\beta){\cal H}_{16}(x_i)-(3+\beta){\cal
H}_{10}(x_i)\Big]+m_N^3\Big[(-8+2t_1-2x_2){\cal
H}_{22}(x_i)\right.\right.\nonumber\\&&\left.\left.+(-1+\beta){\cal
H}_{10,-16}(x_i)-2\beta{\cal
H}_{24}(x_i)\Big]+m_N\Big[Q^2(-1+\beta)(-1+t_1-x_2){\cal
H}_{10,-16}(x_i)\right.\right.\nonumber\\&&\left.\left.+Q^2(-6t_1+6x_2+4+2\beta){\cal
H}_{22}(x_i)+2Q^2\beta(1-t_1+x_2){\cal
H}_{24}(x_i)\Big]\vphantom{\int_0^{x_2}}\right\}+\frac{m_N^3}{M_B^2t_1^2
2\sqrt{2}}\left\{\vphantom{\int_0^{x_2}}{\cal
H}_{6,-18_3,20}(x_i)\right.\right.\nonumber\\&&\left.\left.+(-1+\beta){\cal
H}_{12}(x_i)-{\cal
H}_{6,-13,18}(x_i)\vphantom{\int_0^{x_2}}\right\}+\frac{m_N}{M_B^2t_1
4\sqrt{2}}\left\{\vphantom{\int_0^{x_2}}[Q^2+s(t_1,Q^2)]\Big[(3+25\beta){\cal
H}_{20}(x_i) +2(-1+\beta){\cal
H}_{-6,12}(x_i)\Big]\right.\right.\nonumber\\&&
-\left.\left.(5+\beta){\cal H}_{18}(x_i)-m_N^2\Big[2(-1+\beta){\cal
H}_{-6,12}(x_i) -(11+3\beta){\cal H}_{18}(x_i)+(5+67\beta){\cal
H}_{20}(x_i)\Big]\right.\right.\nonumber\\&&
\left.\left.+2x_2\Big[(-1+\beta){\cal H}_{-10,16}(x_i) +\beta{\cal
H}_{24}(x_i)\Big]-m_Nm_b\Big[{\cal
H}_{6_6,-8_3,-9_3,12_2,14,15,-20_4,21_4}(x_i) +4x_2(-1+\beta){\cal
H}_{22}(x_i)\right.\right.\nonumber\\&& \left.\left.+\beta {\cal
H}_{6_2,-8,-9,12_6,14_3,15_3,20_4,-21_4}(x_i)\Big]\vphantom{\int_0^{x_2}}\right\}
+\frac{m_N}{M_B^24\sqrt{2}}\left\{\vphantom{\int_0^{x_2}}Q^2\Big[{\cal
H}_{-6_2,+12_2,18_9,-20_3}(x_i)+\beta{\cal
H}_{6_2,-12_2,18_3,-20_{47}}(x_i)\Big]\right.\right.\nonumber\\&&
\left.\left.+4m_N(-1+\beta){\cal H}_{22}(x_i) +s(t_1,Q^2)\Big[{\cal
H}_{18_3,-20}(x_i)+\beta{\cal
H}_{18,-20_{21}}(x_i)\Big]\right.\right.\nonumber\\&&\left.\left.
+m_N^2\Big[\beta {\cal
H}_{4_4,8,-9,-10_2,14,-15,16_2,-18,20_{41},-21_4,23_{16},24_4}(x_i)
+{\cal H}_{-2_4,-8,9,10_2,-14,15,-16_2,-18_3,20,-23_4}(x_i)
\right.\right.\nonumber\\&&\left.\left.+8(t_1-x_2){\cal
H}_{22}(x_i)\Big]\vphantom{\int_0^{x_2}}\right\}+\frac{m_N}{t_14\sqrt{2}}\left\{\vphantom{\int_0^{x_2}}2(-1+\beta){\cal
H}_{-6,12}(x_i) +(1+5\beta){\cal H}_{20}(x_i)-(3+\beta){\cal
H}_{18}(x_i)\vphantom{\int_0^{x_2}}\right\}\right.\nonumber\\&&
+\left.\frac{m_N}{4\sqrt{2}}\left\{\vphantom{\int_0^{x_2}}(1+21\beta){\cal
H}_{20}(x_i) -(3+\beta){\cal
H}_{18}(x_i)\vphantom{\int_0^{x_2}}\right\}\vphantom{\int_0^{x_2}}\right]+\int_{t_{0}}^{1}dx_{2}\int_{0}^{1-x_{2}}dx_{1}e^{-s_{0}/M_{B}^{2}}
\left[\vphantom{\int_0^{x_2}}
\frac{m_N^4t_0^2}{(Q^2+m_N^2t_0^2)^3\sqrt{2}}(t_0-x_2)\left\{\vphantom{\int_0^{x_2}}
\right.\right.\nonumber\\&&\left.\left. \left.
\Big[m_N^2m_b(-1+\beta)(-2+t_0)(-1+t_0)
+2m_N^3t_0\{2+(-4+t_0)t_0\}-2m_Nt_0^2\{Q^2(-2+3t_0)+(-2+t_0)s(s_0,Q^2)\}
\right.\right.\right.\nonumber\\&&\left.\left.\left.-m_b(-1+\beta)t_0\{Q^2(-1+3t_0)+(-1+t_0)s(s_0,Q^2)\}\Big]{\cal
H}_{22}(x_i)+m_Nt_0\Big(\{m_N^2(-1+\beta)(-1+t_0)
\right.\right.\right.\nonumber\\&&-\left.\left.\left.m_N
m_bt_0(3+\beta)+(-1+\beta)t_0[Q^2(-1+t_0)-s(s_0,Q^2)]\}{\cal
H}_{10}(x_i)-\Big[m_N^2(-1+\beta)(-1+t_0)+m_Nm_bt_0(1+3\beta)\right.\right.\right.\nonumber\\&&
\left.\left.\left.+(-1+\beta)t_0\{Q^2(-1+t_0)-s(s_0,Q^2)\}\Big]{\cal
H}_{16}(x_i)+2\Big[-m_Nm_bt_0(-1+\beta) +m_N^2\beta(1-t_0)+\beta
t_0(Q^2(1-t_0)\right.\right.\right.\nonumber\\&&
\left.\left.\left.+s(s_0,Q^2))\Big]{\cal
H}_{24}(x_i)\Big)\right.\vphantom{\int_0^{x_2}}\right\}
+\frac{m_N^2}{(Q^2+m_N^2t_0^2)^22\sqrt{2}}(t_0-x_2)\left\{\vphantom{\int_0^{x_2}}\Big[m_N^2m_b(-1+\beta)(-2+t_0)(-1+t_0)
\right.\right.\nonumber\\&&\left.\left.+2m_N^3t_0\{2+(-4+t_0)t_0\}-2m_Nt_0^2\{Q^2(-2+3t_0)+(-2+t_0)s(s_0,Q^2)\}
-m_bt_0(-1+\beta)\{Q^2(-1+3t_0)
\right.\right.\nonumber\\&&\left.\left.+(-1+t_0)s(s_0,Q^2)\}\Big]{\cal
H}_{22}(x_i)+m_Nt_0\Big[\{m_N^2(-1+\beta)(-1+t_0)-m_Nm_bt_0(3+\beta)
+(-1+\beta)t_0[Q^2(-1+t_0)\right.\right.\nonumber\\&&\left.\left.-s(s_0,Q^2)]\}{\cal
H}_{10}(x_i)-\{m_N^2(-1+\beta)(-1+t_0)
+m_Nm_bt_0(1+3\beta)+(-1+\beta)t_0(Q^2(-1+t_0)-s(s_0,Q^2))\}{\cal
H}_{16}(x_i)
\right.\right.\nonumber\\&&\left.\left.+2\{-m_Nm_bt_0(-1+\beta)+m_N^2\beta(1-t_0)
+\beta t_0[Q^2(1-t_0)+s(s_0,Q^2)]\}{\cal
H}_{24}(x_i)\Big]\vphantom{\int_0^{x_2}}\right\} \right.\nonumber\
\end{eqnarray}

\begin{eqnarray}
&&+\frac{m_N}{(Q^2+m_N^2t_0^2)4\sqrt{2}M_B^2t_0}\left.\left\{\vphantom{\int_0^{x_2}}2m_N(t_0-x_2)
\Big[m_N^2m_b(-1+\beta)(-2+t_0)(-1+t_0)
+2m_N^3t_0(2+(-4+t_0)t_0)\right.\right.\nonumber\\&&\left.\left.+m_bt_0(-1+\beta)\{Q^2(1-3t_0)+2M_B^2t_0+(1-t_0)s(s_0,Q^2)\}
+2m_Nt_0^2\{Q^2(2-3t_0)\}+2M_B^2t_0\right.\right.\nonumber\\&&\left.\left.+(2-t_0)s(s_0,Q^2)\Big]{\cal
H}_{22}(x_i)+t_0\Big[m_N^2M_B^2{\cal H}_{6_2,-12_2,-18_6,20_2}(x_i)
+\beta{\cal H}_{-6_2,12_2,-18_2,20_26}(x_i)\Big]+2m_N^4(1-t_0){\cal
H}_{10,16}(x_i)
\right.\right.\nonumber\\&&\left.\left.+m_N^2M_B^2t_0{\cal
H}_{-6_2,12_2,18_{11},-20_5}(x_i) +m_Nm_bM_B^2t_0{\cal
H}_{-6_6,8_3,9_3,-12_2,-14,-15,20_4,-21_4}(x_i) +M_B^2Q^2t_0{\cal
H}_{6_2,-12_2,-18_5,20_3}(x_i)
\right.\right.\nonumber\\&&+\left.\left.m_N^4\beta t_0{\cal
H}_{-10_2,16_2,24_4}(x_i) +m_N^2M_B^2\beta t_0{\cal
H}_{6_2,-12_2,18_3,-20_{67}}(x_i) +m_Nm_bM_B^2\beta t_0{\cal
H}_{-6_2,8,9,-12_6,-14_3,-15_3,-20_4,21_4}(x_i)\right.\right.\nonumber\\&&\left.\left.
+M_B^2Q^2\beta t_0{\cal
H}_{-6_2,12_2,-18,20_{25}}(x_i)+2m_N^4t_0{\cal
H}_{-10,16}(x_i)+m_N^2m_bt_0^2{\cal H}_{-10_6,-16_2,24_4}(x_i)
\right.\right.\nonumber\\&&\left.\left.+m_N^2M_B^2t_0^2{\cal
H}_{-2_4,-8,9,10_2,-14,15,-16_2,-18_3,20,-23_4}(x_i)
+2m_N^2Q^2t_0^2{\cal H}_{10,-16}(x_i)+M_B^2Q^2t_0^2{\cal
H}_{-6_2,12_2,18_9,-20_3}(x_i)
\right.\right.\nonumber\\&&\left.\left.+m_N^4\beta t_0^2{\cal
H}_{10_2,-16_2,-24_4}(x_i)-m_N^3m_b\beta t_0^2{\cal
H}_{10_2,16_6,24_4}(x_i)+m_N^2M_B^2\beta t_0^2{\cal
H}_{4_4,8,-9,-10_2,14,-15,16_2,-18,20_{41},-21_4,23_{16},24_4}(x_i)\right.\right.\nonumber\\&&\left.\left.+m_N^2Q^2\beta
t_0^2{\cal H}_{-10_2,16_2,24_4}(x_i)+M_B^2Q^2\beta t_0^2{\cal
H}_{6_2,-12_2,18_3,-20_{47}}(x_i)-2Q^2t_0^3{\cal
H}_{10,-16}(x_i)+Q^2\beta t_0^3{\cal
H}_{10_2,-16_2,-24_4}(x_i)\right.\right.\nonumber\\&&\left.\left.+M_B^2t_0s(s_0,Q^2){\cal
H}_{6_2,-12_2,-18_5,20_3}(x_i)+M_B^2\beta t_0 s(s_0,Q^2){\cal
H}_{-6_2,12_2,-18,20_{25}}(x_i)+2m_N^2t_0s(s_0,Q^2){\cal
H}_{10,-16}(x_i)\right.\right.\nonumber\\&&\left.\left.+M_B^2t_0^2s(s_0,Q^2){\cal
H}_{18_3,-20}(x_i)+m_N^2\beta t_0^2s(s_0,Q^2){\cal
H}_{-10_2,16_2,24_4}(x_i)+M_B^2\beta t_0^2s(s_0,Q^2){\cal
H}_{18,-20_{21}}(x_i)\right.\right.\nonumber\\&&\left.\left.-2m_N^2x_2\Big[\{(-1+\beta)(-1+t_0)-m_Nm_bt_0(3+\beta)
+(-1+\beta)t_0[-M_B^2-Q^2(1-t_0)+s(s_0,Q^2)]\}{\cal
H}_{10}(x_i)\right.\right.\nonumber\\&&\left.\left.-\Big[m_N^2(-1+\beta)(-1+t_0)+m_Nm_bt_0(1+3\beta)+(-1+\beta)t_0\{-M_B^2-Q^2(1-t_0)
-s(s_0,Q^2)\}\Big]{\cal
H}_{16}(x_i)\right.\right.\nonumber\\&&\left.\left.+2\Big[-m_Nm_bt_0(-1+\beta)+m_N^2\beta(1-t_0)+\beta
t_0\{M_B^2+Q^2(1-t_0)+s(s_0,Q^2)\}\Big]{\cal
H}_{24}(x_i)\Big]\vphantom{\int_0^{x_2}}\right\}\vphantom{\int_0^{x_2}}\right),
\end{eqnarray}
  \begin{eqnarray}\label{f_{2}}
&&f_{2}(Q^{2})= \frac{1}{\sqrt{2}\lambda_{\Sigma_b}}
e^{m_{\Sigma_b}^{2}/M_{B}^{2}}\left(\vphantom{\int_0^{x_2}}\int_{t_{0}}^{1}dx_{2}\int_{0}^{1-x_{2}}dx_{1}
e^{-s(x_{2},Q^{2})/M_{B}^{2}}\frac{1}{2\sqrt{2}x_2}\left[\vphantom{\int_0^{x_2}}{\cal
H}_{11,-17_{2},5}(x_i) -\beta{\cal
H}_{11,-17_{12},5}(x_i)\vphantom{\int_0^{x_2}}\right]
\right.\nonumber\\&&\left.+\int_{t_0}^1dx_2\int_0^{1-x_2}dx_1\int_{t_0}^{x_2}dt_1e^{-s(t_1,Q^2)/M_{B}^{2}
}\left[\vphantom{\int_0^{x_2}}-\frac{m_N^4
}{M_B^4t_1^3\sqrt{2}}(3+\beta)x_2{\cal
H}_{22}(x_i)+\frac{m_N^2}{M_B^4t_1^2
2\sqrt{2}}\left\{\vphantom{\int_0^{x_2}}m_N m_b
x_2\Big[(1+3\beta){\cal
H}_{16}(x_i)\right.\right.\right.\nonumber\\&&
\left.\left.+2(-1+\beta){\cal H}_{24}(x_i)+(3+\beta){\cal
H}_{10}(x_i)\Big]+2\Big[-m_N m_b
x_2(-1+\beta)-\{Q^2+s(t_1,Q^2)\}(3+\beta)x_2+m_N^2(3+\beta\right.\right.\nonumber\\&&\left.\left.+(5+\beta)x_2\Big]{\cal
H}_{22}(x_i)\vphantom{\int_0^{x_2}}\right\} +\frac{m_N^2}{M_B^4 t_1
2\sqrt{2}} \left\{\vphantom{\int_0^{x_2}}m_N m_b
\Big[(1+3\beta){\cal H}_{16}(x_i)+2(-1+\beta){\cal
H}_{24}(x_i)+(3+\beta){\cal
H}_{10}(x_i)\Big]\right.\right.\nonumber\\&&\left.\left.+2\Big[m_N
m_b
(1-\beta)-s(t_1,Q^2)(3+\beta+x_2)+m_N^2(5+\beta+x_2)-Q^2(3+\beta+(4+\beta)x_2)\Big]{\cal
H}_{22}(x_i) \vphantom{\int_0^{x_2}}\right\}
\right.\nonumber\\&&+\left.\frac{m_N^2}{M_B^4\sqrt{2}}\left\{\vphantom{\int_0^{x_2}}\Big[m_N^2-(4+\beta)Q^2-s(t_1,Q^2)\Big]{\cal
H}_{22}(x_i)\vphantom{\int_0^{x_2}}\right\}+\frac{m_N}{M_B^2t_1^2
2\sqrt{2}}\left\{\vphantom{\int_0^{x_2}}-m_b\Big[{n\cal
H}_{6_3,12,-18,-20}(x_i)+\beta{\cal
H}_{6,12_3,18,20}(x_i)\Big]\right.\right.\nonumber\\&&\left.\left.-2
(2+\beta)x_2{\cal
H}_{22}(x_i)\vphantom{\int_0^{x_2}}\right\}+\frac{m_N^2}{M_B^2t_1
4\sqrt{2}}\left\{\vphantom{\int_0^{x_2}}{\cal
H}_{-2_4,-8,9,15,-18_2,-20_2,22_8,-23_4}(x_i) +(-1+\beta){\cal
H}_{14}(x_i)\right.\right.\nonumber\\&& \left.\left.-\beta{\cal
H}_{-4_4,-8,9,15,18_2,-20_{22},21_4,-22_4,-23_{16},}(x_i)-8{\cal
H}_{22}(x_i) \vphantom{\int_0^{x_2}}\right\}
+\frac{m_N^2\sqrt{2}}{M_B^2}{\cal
H}_{22}(x_i)\vphantom{\int_0^{x_2}}\right]\nonumber\\
&&+\left.\int_{t_{0}}^{1}dx_{2}\int_{0}^{1-x_{2}}dx_{1}e^{-s_{0}/M_{B}^{2}}\left[\vphantom{\int_0^{x_2}}
\frac{m_N^4t_0^2}{(Q^2+m_N^2t_0^2)^3\sqrt{2}}(t_0-x_2)\left\{\vphantom{\int_0^{x_2}}
-m_N m_b t_0\Big[(1+3\beta){\cal H}_{16}(x_i)+2(-1+\beta){\cal
H}_{24}(x_i)\right.\right.\right.\nonumber\\&&\left.\left.\left.+(3+\beta){\cal
H}_{10}(x_i)+2(m_N
m_b(-1+\beta)t_0+m_N^2(3+\beta-(5+\beta)t_0+t_0^2)+t_0(Q^2(3+\beta)-(4+\beta)t_0)
\right.\right.\right.\nonumber\\&&\left.\left.\left.+(3+\beta-t_0)s(s_0,Q^2)\Big]{\cal H}_{22}(x_i)\vphantom{\int_0^{x_2}}\right\}
-\frac{m_N^2}{(Q^2+m_N^2t_0^2)^22\sqrt{2}}(t_0-x_2)\left\{\vphantom{\int_0^{x_2}}m_N
m_b t_0\Big[(1+3\beta){\cal H}_{16}(x_i)+2(-1+\beta){\cal
H}_{24}(x_i)\right.\right.\right.\nonumber\\&&\left.\left.+(3+\beta){\cal
H}_{10}(x_i)\Big]+2\Big[-m_N
m_b(-1+\beta)t_0+Q^2t_0\{-3-\beta+(4+\beta)t_0\}+m_N^2\{-3+\beta(-1+t_0)-(-5+t_0)t_0\}\right.\right.\nonumber\\&&
\left.\left.+t_0(-3-\beta+t_0)s(s_0,Q^2)\Big]{\cal
H}_{22}(x_i)\vphantom{\int_0^{x_2}}\right\}
+\frac{m_N}{(Q^2+m_N^2t_0^2)4\sqrt{2}M_B^2t_0}\left\{\vphantom{\int_0^{x_2}}4m_N(t_0-x_2)
\Big[m_N m_b(-1+\beta)t_0 +m_N^2\{3+\beta\right.\right.\nonumber\
\end{eqnarray}
\begin{eqnarray}\label{f_{2}}
&&\left.-(5+\beta)t_0+t_0^2\}+t_0\{M_B^2(2+\beta+2t_0)+Q^2\{3+\beta-(4+\beta)t_0\}
+(3+\beta-t_0)s(s_0,Q^2)\}{\cal
H}_{22}(x_i)\right.\nonumber\\&&\left. -t_0\Big[m_b M_B^2\{{\cal
H}_{6_6,12_2,-20_2}(x_i)+\beta{\cal H}_{6_2,12_6,20_2}(x_i)\}+m_N^2
m_b t_0\{{\cal H}_{10_6,16_2,-24_4}(x_i)+\beta{\cal
H}_{10_2,16_6,24_4}(x_i)\}\right.\nonumber\\&&\left.+m_N
M_B^2t_0\{{\cal H}_{2_4,8,-9,14,-15,20_2,23_4}(x_i)+\beta{\cal
H}_{-4_4,-8,9,-14,15,-20_{22},21_4,-23_{16}}(x_i)\}+2M_B^2\{m_b(-1+\beta)\right.\nonumber\\&&\left.\left.\left.+m_N(1+\beta)t_0\}{\cal
H}_{18}(x_i)-2m_N^2m_b x_2\{(3+\beta){\cal
H}_{10}(x_i)+(1+3\beta){\cal H}_{16}(x_i)+2(-1+\beta){\cal
H}_{24}(x_i)\}\Big]
\vphantom{\int_0^{x_2}}\right\}\vphantom{\int_0^{x_2}}\right]\vphantom{\int_0^{x_2}}\right\},
\end{eqnarray}
where
\begin{eqnarray}
{\cal H}(x_i) &=& {\cal H}(x_1,x_2,1-x_1-x_2),
\nonumber \\
s(y,Q^2)&=&(1-y)m_{N}^2+\frac{(1-y)}{y}Q^2+\frac{m_b^2}{y}.
\end{eqnarray}
The  $t_0=t_{0}(s_{0},Q^2)$ is the solution of the equation
$s(t_{0},Q^2)=s_{0}$, i.e.,
\begin{eqnarray}
t_{0}(s_{0},Q^2)=\frac{m_N^2-Q^2+\sqrt{4m_N^2(m_b^2+Q^2)+(m_N^2-Q^2-s_0)^2}-s_0}{2m_N^2}.
\end{eqnarray}
Here, $s_0$ is continuum threshold, $M_B^2$ is the Borel mass
parameter. In calculations,   the following short hand notations for
the functions ${\cal H}_{\pm i_a,\pm j_b, ...}=\pm a{\cal H}_{i}\pm
b{\cal H}_{j}...$ are used, and  ${\cal H}_{i}$ functions are
written in terms of the DA's in the following way:
\begin{eqnarray}
&&{\cal H}_{1}=S_{1}~~~~~~~~~~~~~~~~~~~~~~~~~~~~~~~~~~~~~~~~{\cal
H}_{2}=S_{1,-2}\nonumber\\&&{\cal
H}_{3}=P_{1}~~~~~~~~~~~~~~~~~~~~~~~~~~~~~~~~~~~~~~~~{\cal
H}_{4}=P_{1,-2}\nonumber\\&&{\cal
H}_{5}=V_{1}~~~~~~~~~~~~~~~~~~~~~~~~~~~~~~~~~~~~~~~~{\cal
H}_{6}=V_{1,-2,-3}\nonumber\\&&{\cal
H}_{7}=V_{3}~~~~~~~~~~~~~~~~~~~~~~~~~~~~~~~~~~~~~~~~{\cal
H}_{8}=-2V_{1,-5}+V_{3,4}\nonumber\\&&{\cal
H}_{9}=V_{4,-3}~~~~~~~~~~~~~~~~~~~~~~~~~~~~~~~~~~~~~{\cal
H}_{10}=-V_{1,-2,-3,-4,-5,6}\nonumber\\&&{\cal
H}_{11}=A_{1}~~~~~~~~~~~~~~~~~~~~~~~~~~~~~~~~~~~~~~~{\cal
H}_{12}=-A_{1,-2,3}\nonumber\\&&{\cal
H}_{13}=A_{3}~~~~~~~~~~~~~~~~~~~~~~~~~~~~~~~~~~~~~~~{\cal
H}_{14}=-2A_{1,-5}-A_{3,4}\nonumber\\&&{\cal
H}_{15}=A_{3,-4}~~~~~~~~~~~~~~~~~~~~~~~~~~~~~~~~~~~~{\cal
H}_{16}=A_{1,-2,3,4,-5,6}\nonumber\\&&{\cal
H}_{17}=T_{1}~~~~~~~~~~~~~~~~~~~~~~~~~~~~~~~~~~~~~~~~{\cal
H}_{18}=T_{1,2}-2T_{3}\nonumber\\&&{\cal
H}_{19}=T_{7}~~~~~~~~~~~~~~~~~~~~~~~~~~~~~~~~~~~~~~~~{\cal
H}_{20}=T_{1,-2}-2T_{7}\nonumber\\&&{\cal
H}_{21}=-T_{1,-5}+2T_{8}~~~~~~~~~~~~~~~~~~~~~~~~~~{\cal
H}_{22}=T_{2,-3,-4,5,7,8}\nonumber \\&&{\cal
H}_{23}=T_{7,-8}~~~~~~~~~~~~~~~~~~~~~~~~~~~~~~~~~~~~~{\cal
H}_{24}=-T_{1,-2,-5,6}+2T_{7,8}, \
\end{eqnarray}
 where for each DA's, we also have used $X_{\pm i,\pm j, ...}=\pm X_{i}\pm
 X_{j}...$.

The explicit expressions for the nucleon DA's is given as:
\begin{eqnarray}\label{DA's}
V_1(x_i,\mu)&=&120x_1x_2x_3[\phi_3^0(\mu)+\phi_3^+(\mu)(1-3x_3)],\nonumber\\
V_2(x_i,\mu)&=&24x_1x_2[\phi_4^0(\mu)+\phi_3^+(\mu)(1-5x_3)],\nonumber\\
V_3(x_i,\mu)&=&12x_3\{\psi_4^0(\mu)(1-x_3)+\psi_4^-(\mu)[x_1^2+x_2^2-x_3(1-x_3)]
\nonumber\\&&+\psi_4^+(\mu)(1-x_3-10x_1x_2)\},\nonumber\\
V_4(x_i,\mu)&=&3\{\psi_5^0(\mu)(1-x_3)+\psi_5^-(\mu)[2x_1x_2-x_3(1-x_3)]
\nonumber\\&&+\psi_5^+(\mu)[1-x_3-2(x_1^2+x_2^2)]\},\nonumber\\
V_5(x_i,\mu)&=&6x_3[\phi_5^0(\mu)+\phi_5^+(\mu)(1-2x_3)],\nonumber\\
V_6(x_i,\mu)&=&2[\phi_6^0(\mu)+\phi_6^+(\mu)(1-3x_3)],\nonumber\\
A_1(x_i,\mu)&=&120x_1x_2x_3\phi_3^-(\mu)(x_2-x_1),\nonumber\\
A_2(x_i,\mu)&=&24x_1x_2\phi_4^-(\mu)(x_2-x_1),\nonumber\\
A_3(x_i,\mu)&=&12x_3(x_2-x_1)\{(\psi_4^0(\mu)+\psi_4^+(\mu))+\psi_4^-(\mu)(1-2x_3)
\},\nonumber\\
A_4(x_i,\mu)&=&3(x_2-x_1)\{-\psi_5^0(\mu)+\psi_5^-(\mu)x_3
+\psi_5^+(\mu)(1-2x_3)\},\nonumber\\
A_5(x_i,\mu)&=&6x_3(x_2-x_1)\phi_5^-(\mu)\nonumber\\
A_6(x_i,\mu)&=&2(x_2-x_1)\phi_6^-(\mu),\nonumber\\
T_1(x_i,\mu)&=&120x_1x_2x_3[\phi_3^0(\mu)+\frac{1}{2}(\phi_3^--\phi_3^+)(\mu)(1-3x_3)
],\nonumber\\
T_2(x_i,\mu)&=&24x_1x_2[\xi_4^0(\mu)+\xi_4^+(\mu)(1-5x_3)],\nonumber\\
T_3(x_i,\mu)&=&6x_3\{(\xi_4^0+\phi_4^0+\psi_4^0)(\mu)(1-x_3)+
(\xi_4^-+\phi_4^--\psi_4^-)(\mu)[x_1^2+x_2^2-x_3(1-x_3)]
\nonumber\\
&&+(\xi_4^++\phi_4^++\psi_4^+)(\mu)(1-x_3-10x_1x_2)\},\nonumber\\
T_4(x_i,\mu)&=&\frac{3}{2}\{(\xi_5^0+\phi_5^0+\psi_5^0)(\mu)(1-x_3)+
(\xi_5^-+\phi_5^--\psi_5^-)(\mu)[2x_1x_2-x_3(1-x_3)]
\nonumber\\
&&+(\xi_5^++\phi_5^++\psi_5^+)(\mu)(1-x_3-2(x_1^2+x_2^2))\},\nonumber\\
T_5(x_i,\mu)&=&6x_3[\xi_5^0(\mu)+\xi_5^+(\mu)(1-2x_3)],\nonumber\\
T_6(x_i,\mu)&=&2[\phi_6^0(\mu)+\frac{1}{2}(\phi_6^--\phi_6^+)(\mu)(1-3x_3)],
\nonumber \\
T_7(x_i,\mu)&=&6x_3\{(-\xi_4^0+\phi_4^0+\psi_4^0)(\mu)(1-x_3)+
(-\xi_4^-+\phi_4^--\psi_4^-)(\mu)[x_1^2+x_2^2-x_3(1-x_3)]
\nonumber\\
&&+(-\xi_4^++\phi_4^++\psi_4^+)(\mu)(1-x_3-10x_1x_2)\},\nonumber\\
T_8(x_i,\mu)&=&\frac{3}{2}\{(-\xi_5^0+\phi_5^0+\psi_5^0)(\mu)(1-x_3)+
(-\xi_5^-+\phi_5^--\psi_5^-)(\mu)[2x_1x_2-x_3(1-x_3)]
\nonumber\\
&&+(-\xi_5^++\phi_5^++\psi_5^+)(\mu)(1-x_3-2(x_1^2+x_2^2))\},\nonumber\\
S_1(x_i,\mu) &=& 6 x_3 (x_2-x_1) \left[ (\xi_4^0 + \phi_4^0 +
\psi_4^0 + \xi_4^+ + \phi_4^+ + \psi_4^+)(\mu) + (\xi_4^- + \phi_4^-
- \psi_4^-)(\mu)(1-2 x_3) \right]
\nonumber \\
S_2(x_i,\mu) &=& \frac{3}{2} (x_2 -x_1) \left[- \left(\psi_5^0 +
\phi_5^0 + \xi_5^0\right)(\mu) + \left(\xi_5^- + \phi_5^- - \psi_5^0
\right)(\mu) x_3 \right. \nonumber \\
 && \left.+\left(\xi_5^+ + \phi_5^+ + \psi_5^0 \right)(\mu) (1- 2
x_3)\right]
\nonumber \\
P_1(x_i,\mu) &=& 6 x_3 (x_2-x_1) \left[ (\xi_4^0 - \phi_4^0 -
\psi_4^0 + \xi_4^+ - \phi_4^+ - \psi_4^+)(\mu) + (\xi_4^- - \phi_4^-
+ \psi_4^-)(\mu)(1-2 x_3) \right]
\nonumber \\
P_2(x_i,\mu) &=& \frac32 (x_2 -x_1) \left[\left(\psi_5^0 + \psi_5^0
- \xi_5^0\right)(\mu) + \left(\xi_5^- - \phi_5^- + \psi_5^0
\right)(\mu) x_3 \right. \nonumber\\
&& \left. + \left(\xi_5^+ - \phi_5^+ - \psi_5^0 \right)(\mu) (1- 2
x_3)\right]\, .
\end{eqnarray}
 The following functions are encountered to the above amplitudes and they can be  defined in
terms of the  eight independent parameters, namely  $f_N$,
$\lambda_1$, $\lambda_2$, $V_1^d$, $A_1^u$, $f_d^1$, $f_d^2$ and
$f_u^1$:
\begin{eqnarray}
\phi_3^0& =& \phi_6^0 = f_N \nonumber \\
 \phi_4^0 &=& \phi_5^0 =
\frac{1}{2} \left(\lambda_1 + f_N\right)
\nonumber \\
\xi_4^0 &=& \xi_5^0 = \frac{1}{6} \lambda_2\nonumber \\
  \psi_4^0
&=& \psi_5^0 = \frac{1}{2}\left(f_N - \lambda_1 \right) \nonumber\\
\phi_3^- &=& \frac{21}{2} A_1^u,\nonumber\\
\phi_3^+ &=& \frac{7}{2} (1 - 3 V_1^d),\nonumber\\
\phi_4^- &=& \frac{5}{4} \left(\lambda_1(1- 2 f_1^d -4 f_1^u) + f_N(
2 A_1^u - 1)\right) \,,
\nonumber \\
\phi_4^+ &=& \frac{1}{4} \left( \lambda_1(3- 10 f_1^d) - f_N( 10
V_1^d - 3)\right)\,,
\nonumber \\
\psi_4^- &=& - \frac{5}{4} \left(\lambda_1(2- 7 f_1^d + f_1^u) +
f_N(A_1^u + 3 V_1^d - 2)\right) \,,
\nonumber \\
\psi_4^+ &=& - \frac{1}{4} \left(\lambda_1 (- 2 + 5 f_1^d + 5 f_1^u)
+ f_N( 2 + 5 A_1^u - 5 V_1^d)\right)\,,
\nonumber \\
\xi_4^- &=& \frac{5}{16} \lambda_2(4- 15 f_2^d)\,,
\nonumber \\
\xi_4^+ &=& \frac{1}{16} \lambda_2 (4- 15 f_2^d)\,,\nonumber\\
\phi_5^- &=& \frac{5}{3} \left(\lambda_1(f_1^d - f_1^u) + f_N( 2
A_1^u - 1)\right) \,,
\nonumber \\
\phi_5^+ &=& - \frac{5}{6} \left(\lambda_1 (4 f_1^d - 1) + f_N( 3 +
4 V_1^d)\right)\,,
\nonumber \\
\psi_5^- &=& \frac{5}{3} \left(\lambda_1 (f_1^d - f_1^u) + f_N( 2 -
A_1^u - 3 V_1^d)\right)\,,
\nonumber \\
\psi_5^+ &=& -\frac{5}{6} \left(\lambda_1 (- 1 + 2 f_1^d +2 f_1^u) +
f_N( 5 + 2 A_1^u -2 V_1^d)\right)\,,
\nonumber \\
\xi_5^- &=& - \frac{5}{4} \lambda_2 f_2^d\,,
\nonumber \\
\xi_5^+ &=&  \frac{5}{36} \lambda_2 (2 - 9 f_2^d)\,,
\nonumber \\
\phi_6^- &=& \phantom{-}\frac{1}{2} \left(\lambda_1 (1- 4 f_1^d - 2
f_1^u) + f_N(1 +  4 A_1^u )\right) \,,
\nonumber \\
\phi_6^+ &=& - \frac{1}{2}\left(\lambda_1  (1 - 2 f_1^d) + f_N ( 4
V_1^d - 1)\right)
\end{eqnarray}

\begin{thebibliography}{99}
\bibitem{susy} G. Buchalla, G. Hiller and G. Isidori, Phys. Rev. D 63 (2000) 014015.
\bibitem{darkmatter} C. Bird, P. Jackson, R. Kowalewski, M. Pospelov, Phys. Rev. Lett. 93
(2004) 201803.
\bibitem{Acosta} D. Acosta et al., (CDF Collaboration), Phys.
Rev. Lett. 96, 202001 (2006).
\bibitem{Aubert1} B. Aubert et
al., (BABAR Collaboration), Phys. Rev. Lett. 97, 232001 (2006);
Phys. Rev. Lett. 99, 062001 (2007); Phys. Rev. D 77, 012002 (2008).
\bibitem{Mattson} M. Mattson et al., (SELEX Collaboration), Phys. Rev. Lett. 89,
112001 (2002).
 \bibitem{Aaltonen1} T. Aaltonen et al., (CDF Collaboration), Phys. Rev. Lett.
99, 052002 (2007); Phys. Rev. Lett. 99, 202001 (2007).
 \bibitem{Chistov} R. Chistov et al., (Belle
Collaboration), Phys. Rev. Lett. 97, 162001 (2006).
\bibitem{Ocherashvili} A. Ocherashvili et al., (SELEX Collaboration), Phys.
Lett. B 628, 18 (2005).
 \bibitem{Abazov1} V.
Abazov et al., (D0 Collaboration), Phys. Rev. Lett. 99, 052001
(2007); Phys. Rev. Lett. 101, 232002 (2008).
\bibitem{Solovieva} E. Solovieva et al., (Belle Collaboration), Phys. Lett. B  672, 1 (2009).
\bibitem{kmayprd} K. Azizi, M. Bayar, A. Ozpineci, Y. Sarac, Phys. Rev. D 80, 036007
(2009).
\bibitem{kmyh} K. Azizi, M. Bayar,  Y. Sarac, H. Sundu, arXiv:0908.1758
[hep-ph].
\bibitem{Lenz}  V. M.  Braun, A. Lenz, M. Wittmann, Phys. Rev. D  73 (2006) 094019.
\bibitem{Gockeler1}  M. Gockeler et al.,  QCDSF Collaboration, PoS LAT2007 (2007) 147, arXiv:0710.2489 [hep-lat].
\bibitem{Gockeler2}  M. Gockeler et al., Phys. Rev. Lett. 101 (2008) 112002,    arXiv:0804.1877 [hep-lat].
\bibitem{QCDSF}  V. M. Braun et al., QCDSF Collaboration, Phys. Rev. D 79, 034504 (2009).
\bibitem{Balitsky} I. I.  Balitsky,  V. M.  Braun,  Nucl.
Phys.     B311 (1989) 541.
\bibitem{17}  V. M.  Braun, A. Lenz, N. Mahnke, E. Stein, Phys.
Rev. D  65 (2002) 074011.
\bibitem{18}  V. M.  Braun, A. Lenz, M. Wittmann, Phys.
Rev. D  73 (2006) 094019;
  A.~Lenz, M.~Wittmann and E.~Stein,
        Phys.\ Lett.\  B  581 (2004) 199.
\bibitem{Braun1b}  V. Braun, R.~J.~Fries, N.~Mahnke and E.~Stein, Nucl. Phys.
B  589 (2000) 381.
\bibitem{8}  V. M.  Braun, A. Lenz, G. Peters, A. V. Radyushkin,
Phys. Rev. D 73 (2006) 034020.
 \bibitem{Mannel} T. Mannel, W. Roberts and Z. Ryzak,
Nucl. Phys. B355 (1991) 38.

\bibitem{alievozpineci}  T. M. Aliev, A. Ozpineci, M. Savci, Phys. Rev. D 65 (2002)
115002.
\bibitem{Chen} C. H. Chen,  C. Q. Geng, Phys. Rev. D 63 (2001) 054005; Phys.
Rev. D 63 (2001) 114024; Phys. Rev. D 64 (2001) 074001.
 \bibitem{ozpineci}  T. M. Aliev, A. Ozpineci, M. Savci, C. Yuce , Phys. Lett. B 542 (2002)
 229.
\bibitem{Ozpineci1} K. Azizi, M. Bayar, A. Ozpineci, Phys. Rev. D 79, 056002
(2009).

\bibitem{Belyaev} V. M. Belyaev,  B. L.  Ioffe, JETP  56 (1982) 493.

\bibitem{Becirevic} D. Becirevic, A. B. Kaidalov , Phys. Lett. B478 (2000) 417.
\bibitem{Ballb} P. Ball, R. Zwicky, Phys. Rev. D71 (2005) 014015.


  \bibitem{Chuan} Chuan-Hung Chen, C.Q. Geng, Phys. Rev. D 64 (2001)
  074001.
\bibitem{CLEO} CLEO Collaboration, G. Crawford et al Phys. Rev. Lett. 75, 624
(1995).
\bibitem{Burasb} A. J. Buras, M. Muenz Phys. Rev. D52 (1995) 186.
\bibitem{Yao:2006px} C. Amsler et al. [Particle Data Group], Phys. Lett. B 667 1 (2008).
\end{thebibliography}
\end{document}